\documentclass[a4paper]{article}
\usepackage{amsmath}
\usepackage{amsfonts}
\usepackage{amssymb}
\usepackage{bbm}
\usepackage[left=15mm, top=15mm, right=15mm, bottom=15mm, nohead]{geometry}
\usepackage{graphicx}
\usepackage{hyperref}

\renewcommand{\[}{\begin{equation}}
\renewcommand{\]
}{\end{equation}}
\renewcommand{\phi}{\varphi}

\title{
  \hfill \normalsize{ITEP--TH--21/24}

  \hfill \normalsize{MIPT--TH--16/24}

  \vspace{0.2cm}

  \LARGE{Towards non--commutative 2--dimensional integrals}
}
\author{P. Suprun\footnote{suprun.pa@phystech.edu}}
\date{%
  \textit{MIPT, Dolgoprudny, 141701, Russia}
}

\begin{document}

\maketitle

\begin{abstract}
  We collect evidence that the notion of path--ordered non--abelian integration admits an extension to two dimensions. We propose the corresponding notion of non--abelian 2--form along the lines of Lie algebroid theory and argue it is an appropriate one. The processes of parallel transport and integration turn out to be subtly different in the 2--dimensional case; we discuss parallel transport along surfaces and present an indirect definition of a non--abelian integral. This integral includes, for specific choices of 2--forms, both abelian integrals and the continuous limit of Baker--Campbell--Hausdorff formula as special cases; it interpolates between those cases and broadly generalizes them, allowing, for example, an analog of path--exponential with local, point--depending commutators to be spoken about. We comment on all these objects, their relations, gauge symmetries and geometrical meaning, and roughly sketch a plausible order--by--order procedure for obtaining formulas for non--abelian integrals. The exposition is reasonably concrete, relying on no notions more abstract than sections of vector bundles and homotopies.
\end{abstract}

\section{Introduction}

It is well--known that invariant theory of integration (i.e. Cartan---de Rham theory) in arbitrary--dimensional manifolds is based on the notion of differential forms: $p$--forms are precisely the objects that can be invariantely integrated along $p$--dimensional contours (or chains). It is also common knowledge that the theory of 1--forms admits a non--abelian (or matrix) generalization, namely the theory of connections on vector bundles. Like 1--forms, connections can be integrated along 1--dimensional contours; the operation is known as path--ordered exponentiation, and the result describes the cumulative effect of small pieces of contour, which are composed in the order dictated by the contour itself. Rather surprisingly, there is currently no analogous theory in higher dimensions: even for the case of dimension 2, there are no objects that could allow us to speak about the cumulative effect of two--dimensional pieces of surfaces, composed non--commutatively according to how these pieces are glued to the complete surface. The previously undertaken attempts, of which we quote \cite{Baez:2005qu} and the theory of gerbes \cite{Breen:2001ie} just to provide an example, unfortunately did not lead to structures described on the level of concreteness necessary for practical work. Other constructions, like \cite{Akhmedov:2005tn,Akhmedov:2005pv}, suffer from internal restrictions \cite{Akhmedov:2005zi}. It would be extremely desirable to have as integratable objects something as down--to--earth as abelian 2--forms $\omega_{\mu\nu}$ and classical 1--dimensional connections $A^\alpha_{\mu \beta}$, which can be easily manipulated quantitatively on the level of formulas\footnote{Since the present paper was first announced as a preprint, Damien Calaque has implicitly suggested to us a beautifully concrete work \cite{yekutieli2015nonabelian} that builds an approach similar to \cite{Baez:2005qu} in a direct Riemann limit--like fashion. Thanks, Damien! Nevertheless, as the approach suggested here is significantly different, we have decided to proceed with it further for its own intrinsic value.}. In this note, a version of such non--abelian 2--forms is proposed; it is, however, not clear whether or not this construction is ``the most general'' one. If we conjecture that this is indeed the case, then, once the theory of 2--dimensional non--abelian integrals is fully developed, any important application of connections should admit an immediate generalization. In particular, the following questions can be asked in advance:

\begin{itemize}
\item In quantum mechanics, in contrast to general QFT, the evolution of a particle moving on a manifold $M$ is described by a connection on the bundle of function spaces on $M$ ($\mathbb{R} \times \mathcal{F}(M)$), called a hamiltonian. Can the evolution of two--dimensional quantum field theory be described by 2--connections on function spaces in an analogous fashion? This would provide a more detailed description of how the global evolution of a quantum system is built of the evolution of small pieces over short periods of time, as well as shed some light on the dependence of path integrals on boundary conditions (i.e., one--dimensional insertions as opposed to operators attached to points). Moreover, when a target space $M$ is compact or even homogeneous, a natural sequence of finite--dimensional truncations of function space is available, in homogeneous case controlled by representation theory. Can a proper approximation scheme for the behavior of 2d QFT by the behavior of these finite--rank truncations be formulated, analogous to finite--number--of--levels approximation in quantum mechanics? That would provide us with a vast generalization of Atiyah--Segal type formalism\cite{Atiyah:1989vu,Segal:2002ei} from topological quantum field theories to generic ones.
\item To any notion of 2--connection, a Yang--Mills type theory can be put into correspondence, with the Lagrangian of the form $\mathcal{L} = (\mathrm{d}A^{(2)})_{\mu\nu\lambda}(\mathrm{d}A^{(2)})^{\mu\nu\lambda}$, where $\mathrm{d}A$ is an exterior differential yet to be defined. What are the properties of such a theory? In particular, for $d=3$: can the famous exact solution tricks of $2d$ Yang--Mills\cite{Cordes:1994fc} be somehow imitated in the case of $3d$ theory of 2--connections?
\item Computation of the index of a superparticle coupled to a gauge field leads to the Atiyah--Singer index theorem\cite{Alvarez-Gaume:1983zxc,Hietamaki:1993bb}. Which kind of index theorem arises from a superstring coupled to a field of 2--connections, and what are the principal corollaries of such a theorem?
\item De Rham theorem\cite{de1931analysis} teaches us that closed $p$--forms on $M$ feel the topology of $M$, namely, their space modulo ``trivial'' exact forms is precisely the $p$--th cohomology space of $M$. In dimension one, this has a non--abelian generalization, which is the theorem that the moduli space of flat connections on $M$ coincides with the representation space of $\pi_1(M)$ in the structure group of the bundle in question\cite{taubes2011differential}. Can the moduli space of 2--connections be described in an analogous fashion in terms of topological invariants of $M$? For example, is it given by representations of some crossed module\cite{brown1999groupoids} of appropriate kind?
\item Using a necessary version of character theory, how to formulate a proper generalization of Chern--Weil theory of characteristic classes\cite{milnor1974characteristic} for 2--connections and their curvatures? Which topological properties of underlying bundles are measured by such classes? They can turn out to be de Rham analogs of \v{C}ech cocycles constructed in \cite{Fiorenza:2010mh}, for example.

\end{itemize}

\section{Motivation and outline}\label{motivation}

The guiding star of the construction described below is an analogy between the properties of one--forms with given differential and flat connections. Both kinds of objects obey differential relations of a kind
\[\label{domegasmth} \partial_\mu \phi^a_\nu - \partial_\nu \phi^a_\mu + \text{(something algebraic)}_{\mu\nu} = 0\]
where (something algebraic)$=\omega_{\mu\nu}$ (a fixed set of functions) for the first case, and $f^a_{bc}\phi^b \phi^c$ for the second case. Such relations are too weak to determine $\phi^a_\mu$, even with initial conditions: for example, in the plane there is only one equation of type \eqref{domegasmth}, but two of the quantities $\phi^a_x$, $\phi^a_y$. However, given $\phi_y^a$, one can restore $\phi_x^a$ uniquely from the initial conditions $\phi_x^a(y=0,x)$. Notice that in that construction, $\phi_y^a$ is given as a function of $(x,y)$, while $\phi_x^a$ is set only on the ``initial interval'' $y=0$ which then ``dynamically moves'' up to higher values of $y$. In this case, $\phi_x^a$ for $y \neq 0$ can no longer be set arbitrarily, but has to be found dynamically from the equation \eqref{domegasmth}:
\[\label{abtrans} \phi_x(x,y) = \phi_x(x,0) + \int\limits_{0}^y \omega_{xy} dY + \partial_x \int\limits_{0}^{y} \phi_y(x,Y) dY\]
for ordinary 1--forms, and
\[\label{conntrans} \phi^a_x(x,y) = Pe^{\int\limits_{0}^{y}\phi^a_y(x,Y)dY}\, \phi^a_x(x,0)\, Pe^{\int\limits_{y}^{0}\phi^a_y(x,Y)dY} + \int\limits_{0}^{y}Pe^{\int\limits_{y}^{Y}\phi^a_y(x,\Upsilon)d\Upsilon}\, (\partial_x \phi^a_{y}(x,Y))\, Pe^{\int\limits_{Y}^{y}\phi^a_y(x,\Upsilon)d\Upsilon}dY\]
for flat connections. Despite these formulas depend on the choice of $x$ and $y$ coordinates, the object $\phi^a_\mu$ which appears as the result of integration, is invariant: if we perform a change of coordinates $(x,y) \mapsto (x',y')$ and find $\phi_{x'}^a$ from scratch using formulas \eqref{abtrans} or \eqref{conntrans}, the result will coincide with $\phi_{x'}^a$ obtained by a change of basis from $(\phi_x^a,\phi_y^a)$. This happens because the initial equation \eqref{domegasmth} is diffeomorphism--invariant and fixes $\phi_x^a$ from $\phi_y^a$ uniquely; therefore, the result in $(x',y')$, being unique, is obliged to coincide with the transform of $(x,y)$. Because of this property, if an interval $I_i$ is set in motion in \emph{arbitrary} fashion, that is, gets included into a continuous sequence of intervals $I_t$, $t \in [0;1]$, one is free to adjust coordinate system so that $y$ plays the role of time variable and use the same formulas to see how this motion acts on 1--forms or connections.

In this work, we suggest generalizing the equation \eqref{domegasmth} to an unrestricted form
\[\label{dABC} \partial_\mu \phi^a_\nu - \partial_\nu \phi^a_\mu + A_{\mu\nu}^a + B_{\mu}^a(\phi_\nu) - B_{\nu}^a(\phi_\mu) + C^a(\phi_\mu,\phi_\nu) = 0\]
where $B$ is linear in $\phi^a$, and $C^a$ is bilinear and skew--symmetric $\phi \times \phi \rightarrow \phi$ (to maintain skew--symmetry in $\mu$ and $\nu$), and all the quantities can depend on a point $(x,y)$. In components,
\[\label{dABCcomp} \partial_\mu \phi^a_\nu - \partial_\nu \phi^a_\mu + A_{\mu\nu}^a + B_{\mu b}^a \phi_\nu^b - B_{\nu b}^a \phi_\mu^b + C^a_{bc} \phi_\mu^b \phi_\nu^c = 0, \]
$C^a_{bc} = - C^a_{cb}$. If only $A$ is present, we return to the case of abelian one--forms. If $C$ tensor contains structure constants of Lie algebra, we return to the flatness condition. We demonstrate that this equation allows one to restore $\phi^a_x$ knowing $\phi^a_y$ and initial conditions without further restrictions on $A$, $B$, $C$ (in particular, without requiring Jacobi identity to hold). It is ensured that invariance property is unconditionally preserved in our construction. On a geometric side, a group of symmetry of this equation, mixing $A$, $B$ and $C$ together, is found, and the geometric structure for which it is an automorphism group is presented. This structure involves so--called \emph{anchored bundles} just like the much--analogous theory of \emph{Lie algebroids}\cite{Severa:2001tze,fernandes2006lectures,meinrenken2017lie}. The principal difference is that we do not require Jacobi identity to hold, as it is conjecturally a form of closedness condition, while we are proposing a sketch of integration theory for closed and non--closed 2--forms alike. In this light, our geometric structure can be called an \emph{almost Lie algebroid}; we do not, however, make any effort to advocate for this terminology. The discussion of the quantities entering this equation and their group of local symmetries occupies section \ref{split}, while the parallel transport is the business of section \ref{sectransport}. The discussion of the geometric picture is postponed to section \ref{geom} not to interfere with the mainline of the exposition.

Now that we have learned how to move intervals, it is a natural idea to apply this knowledge to the processes of shrinking \emph{closed} contours. Imagine we are given a looping interval $I$, $I(0) = I(1)$ that bounds a disk $D^2 \subset \mathbb{R}^2$; can we contract it into a point when a $\phi^a_s$ is hanging over it? Of course, we can construct a sequence of intervals on disk that becomes a point in the limit, but does such a singular homotopy act nicely on $\phi^a_s$? If it did, any (abelian) one--form on the boundary of a disk could be extended to one with a given exterior differential on a bulk, and any classical connection on the boundary could be extended to a flat one. We know for both situations this is not the case: if $d \phi = \omega $ then the Stokes theorem dictates the equality $ \oint \phi = \int\limits_{D} \omega $ which connects the boundary values of $\phi$ with the bulk behavior of $\omega$; analogously, if $Curv(\phi^a) = 0$ then $Pe^{\oint \phi} = \mathbbm{1}$, so only a connection with trivial holonomy can bound a flat connection in the bulk. In both cases, we see that the space of admissible boundary conditions for equations of type \eqref{domegasmth} is given by a single functional (more precisely, a finite set of functionals $f^a[\phi,\omega],\ a = 1, \ldots N$) of its algebraic part and boundary conditions. We also see (especially vividly from the abelian example) that figuring out which boundary conditions on a disk are admissible is equivalent to developing a theory of integration over said disk: both of ``checking functionals'' $\oint \phi - \int \omega$ and $P\exp{\oint \phi^a}$ are global quantities, collecting together the values of $\phi$ and $\omega$ at different points cumulatively\footnote{One could wonder where is $\omega$ in the second formula. The answer is that to define this $P$--exponential one has to give them a group--like structure, that is, to introduce a multiplication between $P\exp(a,b)$ and $P\exp(b,c)$. On the level of $A$ this corresponds to Lie algebra--like structure, that is, to fixing a commutator $[A_x,A_y]$ on top of a ``vector space'' of A's. Of course, the $P$--exponential changes if this Lie algebra structure is introduced in another way in the bulk. This commutator, which enters the equation \eqref{domegasmth} for connections, plays the role of $\omega$ for $P$--exponential. The discussion will become more clear when we come to the continuous limit of Baker--Campbell--Hausdorff formula later in the text.} Because of that, one can hope the proper foundation of two--dimensional integration is intimately connected with such problems of admissible boundary conditions in the general case as well. We discuss the issue of (non-)contractibility of a circle carrying $\phi^a_s$ and propose an order--by--order method for computing an obstacle; however, the computations involved turn out to be too cumbersome to be carried out in the general case. Those troubles keep us busy in sections \ref{ext} and \ref{assault1}, while in the section \ref{assault2} we analyze the expansion of the integral in area (i.e. study the infinitesimal domain) and compute the first few terms explicitly. The section \ref{bch} is devoted to the continuous version of Baker--Campbell--Hausdorff formula\cite{Magnus:1954zz} and its relevance for our inquiries. The results and multiple problems remaining unresolved are summarized in conclusion \ref{conclusion}.

\section{Key players: splittings and 2--forms}\label{split}

To proceed with our investigations further, we are going to need fields of two types: $\phi^a_\mu$ and a ``combined object'' $(A^a_{\mu\nu}, B^a_{\mu b}, C^a_{bc})$. Their roles are analogous to sections of vector bundles and classical connections in one--dimensional theory, respectively: the most important milestone in our construction is a differential equation relating $\phi^a_\mu$ and $A^a_{\mu\nu},B^a_{b\mu},C^a_{bc}$, in which $(A,B,C)$ ``act'' on $\phi$ in a certain way and select a subset of ``covariantly flat'' ones among these. The objects $\phi^a_\mu$ will be dubbed \emph{splittings}\footnote{the rationale for the name will become more clear in the last section} and triples $(A^a_{\mu\nu}, B^a_{\mu b}, C^a_{bc})$ go under the name of \emph{non--abelian 2--forms}.

Locally, a splitting looks like a one--form with values in a vector bundle: $\phi^a_\mu$, $a = 1,\ldots N$, $\mu = 1,\ldots \dim \mathcal{M}$. Here, $\mu$ index is ``spacetime''--like, and $a$ index labels ``intrinsic'' or ``color'' degrees of freedom. Therefore, with basis field and coordinate systems fixed, a splitting is given by $N \times \dim \mathcal{M}$ smooth functions. However, we allow a group of transformations more general than a gauge group of rank--$N$ vector bundle to act on these objects:
\[\label{gauge} \phi_\mu^a \mapsto \alpha_\mu^a + g^a_b \phi_\mu^b \]
In particular, in case of some nontrivial manifold, the transition functions can be of this kind, i.e., linear inhomogeneous, unlike in the classical vector bundle case. Below, we treat as sensible only notions and conditions covariant with respect to transformations \eqref{gauge}. Notice, in particular, that the equality $\phi^a_\mu = 0$ is NOT of such kind: just like classical connections, splittings do not have a ``prescribed origin'' and are prone to shiftings, moving zero away from its place! Because of that, splittings cannot form a vector space; they form an affine space instead.

A non--abelian 2--form is locally given by a triple $(A_{\mu\nu}^a, B^a_{\mu b}, C^a_{b c})$, where both $A$ and $C$ have to be skew--symmetric with respect to lower indices. Despite looking like very different entities, those components are bound together by gauge transforms: under a mapping $\phi^a_\mu \mapsto g^a_b \phi^b_\mu + \alpha_\mu^a $ those components change as
\[\label{gauge2f} \begin{aligned}
  & C^a_{bc} \mapsto (g^{-1})^a_{p}\, g^q_b\, g^r_c\, C^{p}_{qr}\\
  & B^{a}_{\mu b} \mapsto (g^{-1})^a_{p} \left( g^q_b\, B^{p}_{\mu q} + \alpha^q_\mu\, g^r_b\, C^p_{qr} + \partial_\mu g^{p}_b\right)\\
  & A^{a}_{\mu \nu} \mapsto (g^{-1})^a_{p}\, \left( A^p_{\mu\nu} + \alpha^q_\mu\, B^p_{\mu q} - \alpha^q_\nu\, B^p_{\nu q} + \alpha^p_\mu\, \alpha^q_{\nu}\, C^p_{qr} + \partial_\mu \alpha_\nu^p - \partial_\nu \alpha^p_\mu \right) 
\end{aligned} \]
 Notice that while the vanishing of $C^a_{bc}$ part is an invariant fact, the same cannot be said about $B^a_{\mu b}$ and $B^a_{\mu\nu}$: if $C^a_{bc} \neq 0$, their vanishing depends on basis choice, as $\alpha^a_\mu$--terms can mix part of $C^a_{bc}$ into $B^a_{\mu b}$ and $A^{a}_{\mu\nu}$. Like splittings and connections in their usual sense, non--abelian 2--forms are members of an affine space that does not have a predefined vector space structure. To grasp the logic of transformation rules \eqref{gauge2f} a little better, we can introduce a unified array $\omega_{\bullet \bullet}^a$, where $\bullet$ runs over both the range of $\mu$ and range of $a$. Different components of $\omega$ are then $\omega^a_{\mu\nu} := A^a_{\mu\nu}, \omega^a_{b\mu} := - \omega^a_{\mu b} := B^a_{\mu b}, \omega^a_{bc}:= C^a_{bc}$. Under \emph{constant} gauge transformations this object behaves tensorially: 
\[ \omega^a_{i j} \mapsto (g^{-1})^a_b G^k_i G^l_j \omega^b_{k l},\]
 where 
\[ G = \begin{pmatrix} \mathbbm{1}^\mu_\nu & 0^\mu_b \\
 \alpha^a_{\nu} & g^a_b \end{pmatrix}\]
 The complete formula for gauge transformations includes, of course, the derivatives of parameters $\alpha$ and $g$. From now on, even when the convention about the unified array is not held, we will sometimes refer to a 2--form $(A,B,C)$ as $\omega$ for the sake of brevity. In analogous fashion, $\phi^a_\mu$ can be extended to an enlarged matrix $\phi^\bullet_\mu$, where $\phi^\nu_\mu$ part is set to be $\delta^\nu_\mu$--tensor. The block matrix above recovers the transformation rule \eqref{gauge}, for $\phi^a_\mu$ part, while $\delta^\nu_\mu$ part remains unchanged because of the triangular shape of $G$. It is natural to ask, why would one prescribe such complicated--looking rules of transformation to non--abelian two--forms? The answer is that those rules are chosen in such a way that the \emph{$\omega$--exterior derivative} 
\[\label{domega} (\mathrm{d}_\omega \phi)^a_{\mu\nu} = \partial_\mu \phi_\nu^{a} - \partial_\nu \phi_\mu{a} + A_{\mu\nu}^a + B_{\mu b}^a \phi_\nu^b - B_{\nu b}^a \phi_{\mu}^b + C^a_{bc}\phi^b \phi^c \]
is invariant under gauge transformations: the values of this expression in two basis fields, before and after a transform, are related to each other in a covariant fashion: $(\mathrm{d}_{\omega} \phi)^a_{\mu \nu} \mapsto g^a_b (\mathrm{d}_{\omega} \phi)^a_{\mu \nu})^a $. Because of that, it is possible to speak about \emph{$\omega$--flat splittings} -- i.e., ones obeying the equation $(\mathrm{d}_\omega \phi)^a_{\mu\nu} = 0$.

The classical cases discussed above, and the respective equations, are particular cases of this $\omega$--exterior derivative. For abelian one--forms we can consider
\[ A^a_{\mu \nu} = \omega_{\mu \nu},\ B^a_{\mu b} = 0,\ C^a_{bc} = 0\]
where $\omega_{\mu\nu}$ is some ordinary 2--form (recall that the upper index runs over only one value). For classical connections, there is a complementary example:
\[ A^a_{\mu \nu} = 0,\ B^a_{\mu b}=0,\ C^a_{bc} = f^a_{bc}\]
where $f^a_{bc}$ are structure constants of $\mathfrak{gl}(n)$. In other words, this $\omega$ is a commutation part of the curvature formula.

\section{Parallel transport} \label{sectransport}

Before (and somewhat in parallel) we discuss the actual integration of 2--forms, it is necessary to understand another operation commonly associated with connections, namely \emph{parallel transport}. Here is a tiny reflection/recollection about it: ordinary (1--dimensional) connections, being restricted to an interval $[A,B]$, can parallel--transport the vectors of fiber over $A$ to vectors of fiber over $B$. In the process, the vector $v_A^\alpha$ over $A$ becomes a value of a covariantly--constant section over the whole interval. This means it serves as an initial condition to a differential equation $\partial_t v^\alpha - (A_t)^\alpha_\beta v^\beta = 0,\ v(0) = v_A^a$, where $A_{t\beta}^\alpha = A_{\mu\beta}^\alpha (\partial_t x^\mu)$, the transported $v_B^\alpha$ being the value of solution over the endpoint $B$. In this way, a classical connection puts a map $V_A \rightarrow V_B$ into correspondence with any interval joining $A$ and $B$, that is, with any \emph{1--homotopy} $h: [0;1] \rightarrow M,\ h(0)=A,\ h(1) = B$.

Likewise, 2--connections are going to ``move splittings when an interval is moved''. Consider an interval $I_0: [0;1] \rightarrow \mathbb{R}^2$ lying in the plane, and suppose we are given a restriction $\phi_s^a$ on it (here $s$ is the parameter on $I$, $\phi^a_s = (\partial_s x^\mu) \phi_\mu^a$). Imagine then that we wiggle this interval a little bit (with fixed endpoints), so that the point number $s$ gets shifted to $x^\mu(s,t)$ position. This wiggling includes $I_0$ into a continuous family of intervals $I_t$, $t\text{ also } \in [0;1]$. Mathematically speaking, we have just described a \emph{2--homotopy of intervals} $h: I \times I \rightarrow M$, $h(0,\cdot) = \mathrm{const}.$, $h(1,\cdot) = \mathrm{const}.$ which generalize 1--homotopies (paths joining points). Then there is a natural way to transport $\phi^a_s$ from $I_0$ to $I_1$ (notice how functional--sized spaces of splittings replaced vectors over points). In the process, $\phi^a_{s}$ gets spread in $\omega$--flat fashion over the 2--dimensional area swept by $I$ along the way. A subtle nuance is that in the case of 2--homotopies, some additional data is needed to fix the map precisely. Namely, we need to fix the value of extended splitting $\phi$ on the vector field $Y = \partial_t x^\mu(s,t)$ performing the deformation of the interval:
\[\label{transport}\left\{\begin{aligned}
  &\partial_\mu \phi_\nu^{a} - \partial_\nu \phi_\mu^{a} + A_{\mu\nu}^a + B_{\mu b}^a \phi_\nu^b - B_{\nu b}^a \phi_{\mu}^b + C^a_{bc}\phi^b \phi^c = 0\\
  &\phi_\mu^a Y^\mu = \mu^a
\end{aligned}\right.\]
In this equation, $\mu$ is an $N$--component quantity that transforms somewhat analogously to $\phi^a_\mu$: under a gauge transformation $(g^a_b, \alpha^a_\mu)$ $\mu^a$ maps to $g^a_b \mu^b + \alpha^a_\mu Y^\mu$. We see the vector field $Y$ is encoded in the law of transformation for $\mu$, so for different vector fields they are formally quantities of different types. Other than that, their transformation law is a pretty valid representation of the gauge group in question.

For our primary examples, abelian 1--forms and classical flat connections, the first equation of \eqref{transport} is exactly the condition \eqref{domegasmth} we started from. The process of parallel transport coincides precisely with ``determining $\phi_x$ from $\phi_y$ and initial conditions'', which was the first discussed point of section \ref{motivation}. Back then, the role of time was played by the transversal coordinate $y$.

The analysis of the first equation of system \eqref{transport} can be done the most easily if $Y$ is straightened by a coordinate transformation, so the initial interval is $[0,1] \times \{0\}$, and $Y = \begin{pmatrix} 0 \\ f(x) \end{pmatrix}$ where $f(0) = f(1) = 0 $ (so $Y$ is aligned along the $y$ axis). In this case, with $\mu$ fixed, \eqref{transport}.1 becomes a first--order linear inhomogeneous ODE in $\phi_x$. It is well--known that a Cauchy problem for equations of this type always possesses a solution; therefore, given $\phi^a_x(x,0)$, $\phi^a_x(x,t f(x))$ is, indeed, uniquely determined. As all the coefficients and initial conditions for the equation are smooth with respect to $x$, we can take an arbitrary number of derivatives of $\phi^a_x(x, t f(x))$; therefore, the transported splitting is also smooth.

Just like the initial inspirative formulas \eqref{abtrans}, \eqref{conntrans}, this parallel transport process is invariant by construction, in the following sense: if we perform a change of coordinates $(x,y) \mapsto (x',y')$ and find $\phi_{x'}^a$ by solving the equation \eqref{transport} from scratch, this time using new $\phi_{y'}^a$ as transporter data, the result will coincide with $\phi_{x'}^a$ expressed as a linear combination of $(\phi_x^a,\phi_y^a)$ by a change of basis in $\phi^a_\mu dx^\mu$. This happens because the initial equation \eqref{transport} is diffeomorphism--invariant and fixes $\phi_x^a$ from $\phi_y^a$ uniquely; therefore, the result of integration in $(x',y')$ coordinates, being unique, has to coincide with the transform of the result in $(x,y)$.

It could seem weird that our ``representation of homotopies'' is not fixed uniquely by only the equations of type \eqref{domega}, but depends on additional data $\phi^a_t$; however, because of these additional data the resulting map can be invariant in the above--described sense: the resulting mappings commute with diffeomorphisms $(x,y) \mapsto (x',y')$ and therefore essentially do not depend on the choice of homotopy $I_0 \dashrightarrow I_1$, but only on the area swept along the way. One can say the equations of type \eqref{domega} capture only the invariant part of homotopy that depends on the area, but not on the choice of $(s,t)$ on the film joining $I_0$ and $I_1$.

To be further comforted about the additional data involved in parallel transport, notice that this ambiguity is present even for ordinary, abelian 2--forms. If you know a one--form $\phi$ on a half of the boundary of a disk and you ensure that $\mathrm{d}\phi = \omega$, this does not tell you much about $\phi$ on the second half! You can move the contour and add to $\phi$ value on (a vector tangent to) each small piece of contour an integral of $\omega$ over the area swept by this piece along the way, but there are more than one way to deform one half of a circle into another half. Not only does the resulting form depend on the mapping between two intervals realized by homotopy (it sure does) --- if the form $\omega$ is non--constant inside the disk, the $\phi$ on the target coast is going to depend on how the contour was moved along the way (for example, if there is a ``bulk'' region inside the disk where $\omega$ has big coefficients, it'll surely be important which part of the interval passed through this bulk, etc.). Only the integral of $\phi$ along the remaining part of a boundary is fixed by the condition $\mathrm{d}\phi = \omega$. Analogously, if we are given a connection on half of the boundary of a disk and we try to extend it to another half of said boundary, we cannot detect when a gauge transform localized near the opposite coast is performed, so there is no invariant (canonical) way to get rid of $\mu$ data.

\section{Extension problem and the definition of integrals}\label{ext}

In this section, we move from 2--dimensional parallel transport to 2--dimensional integration. We focus on the case of integration over a disk (or any other contractible domain), which is the simplest one and at the same time the most fundamental, as all the other contours can be glued from disks. The definition of integrals is related to the solvability/unsolvability of a special differential problem regarding non--abelian 2--forms, \emph{the extension problem}. In the case of integration over a disk, this problem is essentially parallel transport from a closed contour to a point, which makes it different from ordinary transport between two contours. Because such contractions to points are singular operations, they often do not act smoothly on spllittings; essentially, non--abelian integrals measure the failure of contractions to define nice parallel transport. A more precise formulation is the following.

To pose an extension problem for a non--abelian 2--form $\omega = (A^a_{\mu\nu},B^a_{\mu b}, C^a_{bc})$, one has to fix an inital value of a splitting $\phi^a_\mu = \phi_\mu^{a(0)}$ on the boundary $\partial D$ of said disk. This means, first, that outside this boundary $\phi^{(0)}$ is not defined, and, second, only the component of $\phi^{(0)}$ ``along the boundary'' is given. To be clear: if $D$ is a standard disk on the plane with Cartesian coordinates $x$, $y$, then only a combination $ - y \phi_x^{a(0)} + x \phi_y^{a(0)}$ in the point $(x,y) \in \partial D$ is initially set, but NOT $\phi_x^{(0)}$ and $\phi_y^{(0)}$ separately. This combination corresponds to a contraction with a vector $\begin{pmatrix} -y \\ x \end{pmatrix}$ tangent to the boundary. With these initial data, the extension problem asks the following question: does there exist a \emph{smooth} $\widetilde{\phi}_\mu^{a}$ defined on the whole disk that coincides with $\phi^{(0)}$ on the boundary and obeys the $\omega$--flatness equation
\[\label{domegaext} (\mathrm{d}_\omega \widetilde{\phi})^a_{\mu\nu} = \partial_\mu \widetilde{\phi}_\nu^{a} - \partial_\nu \widetilde{\phi}_\mu^{a} + A_{\mu\nu}^a + B_{\mu b}^a \widetilde{\phi}_\nu^b - B_{\nu b}^a \widetilde{\phi}_{\mu}^b + C^a_{bc}\widetilde{\phi}^b_\mu \widetilde{\phi}^c_\nu = 0 \]
everywhere? Notice that, as there are only two coordinates on a disk, it is safe to replace $\mu$ with $x$ and $\nu$ with $y$ in this equation.

To comprehend the problem better and to see that the existence of $\widetilde{\phi}^a_\mu$ is not tautological and requires conditions, let us consider classical examples first. In those cases, we ask
\begin{itemize}
\item when an abelian 1--form $\phi^{(0)}_\theta$ on a circle can be extended to $\widetilde{\phi}$ on the whole disk with $\partial_\mu \widetilde{\phi}_\nu - \partial_\nu \widetilde{\phi}_\mu + \omega_{\mu\nu} = 0$?
\item when a classical connection $\phi^{a(0)}_\theta$ can be extended to $\widetilde{\phi}^a_\mu$ on the whole disk with $\partial_\mu \widetilde{\phi}^a_\nu - \partial_\nu \widetilde{\phi}^a_\mu + [\widetilde{\phi}_\mu,\widetilde{\phi}_\nu]^a = 0$?
\end{itemize}
We know that in those cases, bulk equations put restrictions on boundary values: one can determine the integral of one--form from its exterior differential, and flat connections have trivial holonomy over every loop. In other words, the answers to those questions are
\begin{itemize}
\item  if and only if $\oint \phi^{(0)}_\theta\, \mathrm{d}\theta + \int_{D} \omega_{r\theta}\, r \mathrm{d}r \wedge \mathrm{d}\theta = 0$\\and
\item if and only if $\log P\exp \oint \phi^{a(0)}_\theta\, \mathrm{d}\theta = 0$
\end{itemize}
respectively. So both times we see that the existence of $\widetilde{\phi}_\mu^a$ for given boundary conditions is not trivially granted, but controlled by a single (yet multicomponent) functional of these boundary conditions: one has to require $\oint \phi^{(0)}_\mu \mathrm{d}x^\mu + \int \omega_{\mu\nu} \mathrm{d}x^\mu\wedge\mathrm{d}x^\nu = 0$ for abelian one--forms, and $\log P\exp \oint \phi^{(0)}_\mu \mathrm{d}x^\mu = 0$ for classical connections. Only under these conditions does the solution to the extension problem exist.

This leads us to hypothesize that for a general 2--form $\omega = (A_{\mu\nu}^a,B^a_{\mu b},C^a_{bc})$ the solvability of the extension problem with boundary splitting $\phi^{a(0)}_{\partial D}$ is also controlled by a single $N$--component functional in $\phi^{(0)}$. We call this functional $I^a_{\omega}[\phi^{a(0)}_\mu]$ \textit{the non--abelian integral of $\omega$}. We want it to have the following property:
\[ I^a_{\omega}[\phi^{a(0)}_\mu] = 0 \text{ if and only if equation \eqref{domegaext} has a solution with }\phi^{(0)}\text{ as boundary condition} \]
This line serves as (the best we currently have as) a definition.

Maybe the essence of the problem of integration and its relation to parallel transport is more clear in polar coordinates $r,\theta$. In these terms, we are given only $\phi^{a(0)}_\theta(r=1,\theta)$ initially, while $r$--component is at our disposal. Once we come up with a $\widetilde{\phi}_r^{b}$, we can define $\widetilde{\phi}_\theta^{a}$ on the whole disk by solving the equation \eqref{domegaext} with $\mu = r$, $\nu = \theta$ for $\widetilde{\phi}_\theta^{a}$ with $\phi_\theta^{a(0)}$ as initial conditions --- that is, by parallel transport of $\phi^{a(0)}_\theta$ towards the center with $\widetilde{\phi}_r^a$ as transporter data. Our goal is to tweak $\phi_r$ in such a way that the quantities
\[\begin{aligned}
\widetilde{\phi}_x^a = \cos(\theta) \widetilde{\phi}_r + \frac{\sin(\theta)}{r} \widetilde{\phi}_\theta \\
\widetilde{\phi}_y^a = \sin(\theta) \widetilde{\phi}_r - \frac{\cos(\theta)}{r} \widetilde{\phi}_\theta
\end{aligned}\]
are smooth everywhere, even at $r = 0$. It is more or less clear that when such a $\widetilde{\phi}_r$ exists, it is not unique: for example, in the case of classical connections one can apply a compactly--supported gauge transform to both $\widetilde{\phi}_r$ and $\widetilde{\phi}_\theta$, preserving the relation \eqref{domegaext} between them. However, for some $\phi^{(0)}$ such a $\widetilde{\phi}_r$ does not exist at all. For example, if $\omega = 0$ and $\oint\limits_{\partial D = S^1}\phi_\theta^{(0)} \neq 0$ then no $\widetilde{\phi}_r$ can force $\partial_r \widetilde{\phi}_\theta - \partial_\theta \widetilde{\phi}_r = 0$ to hold everywhere (as it would contradict Stokes theorem). Such $\phi^{(0)}_\theta$ can be extended, using parallel transport, to some annulus inside the disk, but they don't ``glue up well'' in the center of the disk. We want to be able to tell, given $\phi^{a(0)}_\theta$, whether or not an adapted $\widetilde{\phi}^a_r$ exists for it by evaluating our $I_\omega^a$ functional on $\phi^{a(0)}_\theta$ and seeing if it vanishes.

Classical examples considered above teach us that for a single--component $\omega_{\mu\nu}$, $I_{\omega}[\phi^{(0)}] = \oint \phi^{(0)}_\mu \mathrm{d}x^\mu + \int \omega_{\mu\nu} \mathrm{d}x^\mu\wedge\mathrm{d}x^\nu$, and for $\omega$ consisting only of tensor part $\omega^a_{bc}$ which are constant and are structure constants of matrix commutator, $I_\omega^a[\phi^{(0)}] = \log P\exp \oint \phi^{(0)}_\mu \mathrm{d}x^\mu $. These formulas should have a generalization for arbitrary $\omega$'s, for example, as a series over $\omega$.

\section{Comparison with 1--dimensional integrals}

In one dimension, the analogous problem would be like the following. Suppose we are doing integration over an interval, say $[0,1]$. Then we need to be given the following data:
\begin{itemize}
\item On the whole interval there should be a matrix--valued 1--form $A_{\mu a}^b$ (in this case, index $\mu$ runs only over one value).
\item On the ends of this interval we need to be given two vectors (in the space on which $A$ can act): $u^a$ over the point $0$ and $v^b$ over the point $1$
\end{itemize}
Then we'd like to solve the following problem: does there exist a smooth vector $\chi^a$ defined on the whole interval such that $\chi(0)^a = u^a$, $\chi(1)^b = v^b$ and
\[\label{dA} \partial_\mu \chi^a + A_{\mu b}^a \chi^b = 0\]
everywhere on the interval?

We know that not for every pair $(u,v)$ such a solution exists. The possibility or impossibility of solving this problem depends on single, $N$--component, $A$--dependent quantity:
\[ I^a_{\dim =1 } = v^a - \left(P\exp \int A \right)^a_b u^b\]
Here, $P\exp$ is understood in the classical sense. Of course, computing this $I^a_{\dim = 1}$ as a function of $u$,$v$ and $A$ would be equivalent to inventing the $P$--exponent, if we did not know it already.

We can even make a little table comparing the one--dimensional case with the two--dimensional one:

\begin{center}\begin{tabular}{c|c|c} & Dimension 1 & Dimension 2 \\
 \hline Initial data on a bulk & $A_{\mu b}^a$ & $\omega_{\mu\nu}^a$, $\omega_{\mu b}^a$, $\omega^a_{bc}$ \\
 \hline Initial data on a boundary & $u^a(0)$, $v^b(1)$ & $\phi^{a(0)}_\mu\left|_\text{boundary}\right.$ \\
 \hline Extension of initial data & $\chi^a(t)$ & $\widetilde{\phi}^{a}_\mu (x,y)$ \\
 \hline Local condition on this extension & formula \eqref{dA} & formula \eqref{domegaext} \\
 \hline Obstruction to extension & $I^a_{\dim =1} = v^a - (P\exp\int A)^a_b u^b$ & $I^a_\omega [\phi^{(0)}]$ \\
 \hline\end{tabular}

\end{center}

Notice, however, an important difference between the integrals being discussed and their one--dimensional forerunners. While $P$--exponential of a classical connection contains a finite amount of data (it is, basis fixed, a certain $n \times n$ matrix, so it belongs to a finite dimensional space), even our classical examples are functionals: for abelian one--forms, we have a formula $\oint \phi - \int \omega$, which depends on the infinitely--dimensional variable $\phi$ even for $\omega$ fixed, and for classical connections, the corresponding object is $\log (Pe^{\oint \phi^a})$, which is again to be thought of as a function of $\phi$. This can seem puzzling at first, but is actually a situation to be expected: even the first attempt to define an area-ordered integral by considering a triangulation over a disk and replacing cells with tensors and their gluings with tensor contraction forces us, when passing to the continuum limit, to consider triangulations with arbitrarily many edges on the boundary. Such tensor spaces have unboundedly large dimensions, and this feature leads to functional--size freedom for continuum objects. This phenomenon affects the general construction as well: our integrals are functionals rather than finite--dimensional matrices or tensors.

\section{Assault on formula for obstacle 1: Fourier coefficient matching}\label{assault1}

So how can we tell if the equation \eqref{domegaext}, with given $\omega$ and boundary condition $\phi_\theta^a (r=1,\theta) = \phi_\theta^{a(0)}(\theta)$ allows for a choice of $\widetilde{\phi}_r$ such that the resulting object $(\widetilde{\phi}_r^a,\widetilde{\phi}_\theta^b)$ is smooth? Before we dig deep into a computational attempt to answer this question, it is useful to put the following idea into mind. Suppose we stubbornly try to contract a contour with a generic $\phi^a_\theta$ on the boundary --- which problem will we face? The answer is that when the length of the remaining piece of contour is small of order $\epsilon$, the coefficients of our form will grow at a rate of $\sim 1/\epsilon$. This growth comes from our attempt to stuff the integral $\oint\phi$ over an initially--long contour into a small $\epsilon$--sized interval (in classical cases, the ``remaining'' integral is $\oint \phi - \int \omega$ for abelian forms, and the holonomy integral $P e^{\oint \phi^a}$ for classical connections). As the integral (with $\omega$ correction when necessary) over a moving contour does not change along the way, at the size $\epsilon$ the form over it $\Phi_\epsilon$ should have the necessary value of integral over small arc, that is $\int\limits_0^\epsilon \Phi_\epsilon = \oint \phi - \int \omega \approx \Phi_\epsilon \epsilon$ in abelian case, $P\exp\left\{\int\limits_0^\epsilon \Phi_\epsilon \right\} = Pe^{\oint \phi^a} \approx \exp{\epsilon \Phi_\epsilon}$ for connections, so $\Phi_\epsilon = \frac{1}{\epsilon} (\oint \phi - \int \omega )$ or $\Phi_\epsilon = \frac{1}{\epsilon}\log (Pe^{\oint \phi^a})$ respectively. We observe that, unless the numerators of these formulas vanish, $\Phi_\epsilon$ cannot be smooth as $\epsilon \rightarrow 0$. Therefore, an obstacle to contractibility can be thought of as a sort of ``residue'' with respect to $\epsilon$ when making an attempt at contraction. This residue should restore the expressions $(\oint \phi - \int \omega )$ or $\log (Pe^{\oint \phi^a})$ when applied to respective particular cases. While it might look very differently, what goes below is basically a massively evolved attempt to expand on this idea.

As we have to deal with a differential problem on a disk, a natural decision would be to work in polar coordinates, which allow us to use the rotational symmetry of the disk\footnote{Using the full \emph{diffeomorphism} group of the disk, however, seems to require way more moral strength than we have}. Therefore, we need to have a good criterion to tell whether given functions $(\phi_r^a,\phi_\theta^a)$ can represent a $\phi$ that is smooth even at zero --- a point, strictly speaking, not covered by polar coordinates. To nail down this problem, we'd better start with an analogous question about smooth functions $f(r,\theta)$ as opposed to form--like $\phi$'s. For functions, there is a simple criterion:\\
\begin{center} A smooth function $f(r,\theta) = \sum f_n(r) e^{i n \theta}$ in polar coordinates has its $n$'th Fourier component $f_n(r)$ vanishing as $r^{|n|}$ near $r=0$\end{center}This is an easy consequence of $f(x,y)$ Taylor expansion: $f(x,y) = a_{0,0} + a_{1,0} x + a_{0,1} y + \ldots + \sum\limits_{k=0}^{n} a_{k,n-k} x^k y^{n-k} + o(r^n) $. Notice that a term $x^k y^l$ can contribute at most to $k + l$'th Fourier component. Moreover, as a degree--$n$ homogeneous polynomial cannot contain Fourier harmonics of parity different from $n$, the Taylor expansion with respect to $r$ for $f_n (r)$ must have the form $r^{|n|} (f_{n;0} + f_{n;1} r^2 + f_{n;2} r^{4} + \ldots)$ i.e., go with the step of $2$. So a combined Fourier--Taylor expansion of a smooth function $f(r,\theta)$ must be
\[ f(r,\theta) = \sum\limits_{n \in \mathbb{Z}\, k \geqslant 0} f_{n;k} r^{|n|+2 k} e^{i n \theta}\]
The reality condition is $f_{-n;k} = \overline{f_{n;k}}$

Now, we know that a 1--form $\widetilde{\phi}$ on a disk is smooth if and only if the functions $\widetilde{\phi}_x$, $\widetilde{\phi}_y$ for everywhere--defined (Cartesian) coordinates $x,y$ are smooth. Therefore, from the formulas
\[\begin{aligned}
\label{xy2rtheta}\widetilde{\phi}_r = \frac{1}{r} (x \widetilde{\phi}_x + y \widetilde{\phi}_y)\\
\widetilde{\phi}_\theta = -y \widetilde{\phi}_x + x \widetilde{\phi}_y
\end{aligned}\]
and the above--described conditions of smoothness for $f_x$, $f_y$, we obtain a general form of $\widetilde{\phi}_r$, $\widetilde{\phi}_\theta$ expansions for $\widetilde{\phi}$ smooth:
\[\begin{aligned}
\widetilde{\phi}_r = \sum\limits_{n \in \mathbb{Z}\, k \geqslant 0} (\widetilde{\phi}_r)_{n;k} r^{|n|+2k-1} e^{i n \theta}\\
\widetilde{\phi}_\theta = \sum\limits_{n \in \mathbb{Z}\, k \geqslant 0} (\widetilde{\phi}_\theta)_{n,k} r^{|n|+2k} e^{i n \theta}
\end{aligned}\]
with an additional condition $(\widetilde{\phi}_r)_{0,0} = (\widetilde{\phi}_\theta)_{0,0} = 0$ (notice from \eqref{xy2rtheta} that because of additional multiplication on $x,y$ no regular component of $\widetilde{\phi}_x$ or $\widetilde{\phi}_y$ can give rise to $(0,0)$th parts of $\widetilde{\phi}_r$, $\widetilde{\phi}_\theta$). For a 2--form--like quantity $A_{r\theta}$ we obtain (cf. $\mathrm{d} x \wedge \mathrm{d} y = r \mathrm{d}r \wedge \mathrm{d} \theta$)
\[ A_{r\theta} (r,\theta)= r A_{xy} = \sum\limits_{n \in \mathbb{Z}\, k \geqslant 0} (A_{r\theta})_{n;k} r^{|n| + 2k +1} e^{i n \theta}\]
The generalization for multicomponent quantities $\widetilde{\phi}_r^a$, $\widetilde{\phi}_\theta^a$ etc. is almost trivial: the only change is that Fourier--Taylor coefficients are now multicomponent as well:
\[ \widetilde{\phi}_{r}^a = \sum\limits_{n \in \mathbb{Z}\, k \geqslant 0} (\widetilde{\phi}_r^a)_{n;k} r^{|n|+2k-1} e^{i n \theta} \]

\[\label{phitheta} \widetilde{\phi}_{\theta}^a = \sum\limits_{n \in \mathbb{Z}\, k \geqslant} (\widetilde{\phi}_\theta^a)_{n;k} r^{|n|+2k} e^{i n \theta} \]

\[ C^a_{bc} = \sum\limits_{n \in \mathbb{Z}\, k \geqslant 0} (C^a_{bc})_{n;k} r^{|n|+2k} e^{i n \theta} \]

\[ B^a_{r b} = \sum\limits_{n \in \mathbb{Z}\, k \geqslant 0} (B^a_{r b})_{n;k} r^{|n|+2k-1} e^{i n \theta}\]

\[B^a_{\theta b} = \sum\limits_{n \in \mathbb{Z}\, k \geqslant 0} (B^a_{\theta b})_{n;k} r^{|n|+2k} e^{i n \theta}\]

\[A^a_{r\theta} = \sum\limits_{n \in \mathbb{Z}\, k \geqslant 0} (A^a_{r\theta})_{n;k} r^{|n|+2k+1} e^{i n \theta}\]
where $(B^a_{r b})_{0;0} = (B^a_{\theta b})_{0;0} = 0$, $(\widetilde{\phi}^a_{r})_{0;0} = 0$. Notice that we do not require $(\widetilde{\phi}_\theta^a)_{0;0} = 0$ as this $\widetilde{\phi}_\theta$ is going to be dynamically generated from the differential equation \eqref{domegaext} and can, for general initial conditions, be non--smooth.

Which initial conditions are we given in these terms? We know only the values of $\phi_\theta$ for $r=1$, but as a function of $\theta$. This means that arbitrary Fourier components of $\phi_\theta^{a(0)}$ can be determined:
\[\label{bdry} (\phi_\theta^{a(0)})_n = \sum\limits_{k \geqslant 0} (\phi_{r}^a)_{n;k} \]
Therefore, everything we know (initially) about the quantities $(\widetilde{\phi}_{\theta}^a)_{n;k}$ from \eqref{phitheta} are their $k$--wise sums.

Summing up, we can formulate the extension problem in terms of Fourier--Taylor coefficients the following way. Initially we are given the quantities $(A,B,C)^a_{n;k}$ and $(\phi^{a(0)}_\theta)_{n}$ and have to come up with another ones $(\widetilde{\phi}^a_\theta)_{n;k}$, $(\widetilde{\phi}^a_r)_{n;k}$ such that $(\widetilde{\phi}^a_\theta)_{0;0} = 0$, $(\widetilde{\phi}^a_r)_{0;0} = 0$, and equations \eqref{bdry}, \eqref{domegaext} hold (for every $r$ and $\theta$). Of course, these sought components are not defined uniquely by the initial conditions: we have two $(n,k)$--sets of multicomponent quantities, but only one all--$r$, all--$\theta$ equation. We therefore have to come up with an ansatz for $(\widetilde{\phi}_r^a)_{n;k}$ to make our calculations more predictable. To make a good guess, let us write $n$'th Fourier component of the equation \eqref{domegaext} explicitly:
\[\begin{aligned}
\label{domegaFourier}& \sum (|n|+2k) (\widetilde{\phi}_\theta^a)_{n;k} r^{|n|+2k-1} - i n \sum (\widetilde{\phi}_r^a)_{n;k} r^{|n|+2k-1} + \sum (A_{r\theta}^a)_{n;k} r^{|n|+2k+1} +\\
& + \sum (B_{r b}^a)_{n-m;k} (\widetilde{\phi}_{\theta}^b)_{m;l} r^{|m|+|n-m| + 2(k+l) -1} - \sum (B_{\theta b}^a)_{n-m;k} (\widetilde{\phi}_{r}^b)_{m;l} r^{|n-m|+|m| + 2(k+l)-1} + \\
& + \sum (C_{bc}^a)_{n-k-l;s} (\widetilde{\phi}_r^b)_{k;t} (\widetilde{\phi}_\theta^c)_{l;u} r^{|n-k-l| + |k| + |l| + 2(s+t+u)-1} = 0
\end{aligned}\]
We observe that to the $(n,k)$'th component of this equation only the components with $k' < k$ contribute, which allows us to solve it order--by--order in $k$, with lower--$k$ components being fixed when it comes to determining higher--$k$ components. For $n \neq 0$, the coefficient of $(\widetilde{\phi}_{r}^a)_{n;k}$ in $(n;k)$'th equation is $-i n + ((A,B,C)\text{-dependent})$, so we can just solve this equation for it with any ansatz for $\widetilde{\phi}_\theta$: in any case, the necessary $(\widetilde{\phi}_{r}^a)_{n;k}$ is going to be of the form $\frac{i}{n} (1 - \frac{i}{n}((A,B,C)\text{-terms}))^{-1\;a}_b (\text{something})^b$, which admits a good Taylor expansion for $A,B,C \rightarrow 0$. The most naive ansatz we can pick is to pack all of $(\phi_{\theta}^{a(0)})_{n}$ into the first possible order in $r$:
\[\begin{aligned}
(\widetilde{\phi}_\theta^{a})_{n;0} = (\phi_\theta^{a(0)})_n\\
(\widetilde{\phi}_\theta^{a})_{n;k > 0} = 0
\end{aligned}\]
As this requires $(\widetilde{\phi}_\theta^a)_n$ to decrease at the slowest possible rate when $r \rightarrow 0$, such an ansatz should put the least burden on $\widetilde{\phi}_r$. The equation \eqref{domegaFourier} can then be satisfied for $n \neq 0$ with the help of $(\widetilde{\phi}_r^a)_{n;k}$ for each $k$.

However, this maneuver is not applicable for making $n=0$ components of \eqref{domegaFourier} hold. In these components, the coefficients of $(\phi_r^b)_{0;k}$ have the form $ - \sum (C_{bc}^a)_{-l;s} (\phi_\theta^b)_{l;k-|l|-s}$, i.e., they vanish as $C \rightarrow 0$. Therefore, if we tried to make the $(0;k)$'th equation hold with $(\widetilde{\phi}_r^b)$'s, as we did for $(n \neq 0; k)$'th equations, then the expressions for $(\widetilde{\phi}_{r}^a)_{0;k}$ would contain $C^{-1}$ and so be singular at $C = 0$. As our approach is basically perturbative over $(A,B,C)$, this situation is unacceptable. To fix it, notice that the coefficient before $(\widetilde{\phi}_\theta)^a_{0;k}$ in $(0,k)$'th equation is $2k$, and therefore we \emph{can} nonsingularly tweak $\widetilde{\phi}_r^a$ to make the equation \eqref{domegaFourier} satisfied. This, however, leaves a problem: we have both $(\widetilde{\phi}_r^a)_{0;k}$ and $(\widetilde{\phi}_\theta^a)_{0;k}$ as free variables in $(0;k)$'th equation, but only one ($r$--component) equation in $r^{2k-1}$'th Taylor coefficient of $0$'th Fourier harmonic. We therefore have to pick an ansatz for $(\widetilde{\phi}_r)^a_{0,k}$ this time, and we could not think of anything better than $(\widetilde{\phi}_r)^a_{0,k} = 0\ \forall k $. So the final strategy for attacking the extension problem can be formulated the following way: given $(\phi_\theta^{a(0)})_n$ for every $n$, and all components of $A^a_{r\theta}(r,\theta)$, $B_{rb}^a (r,\theta)$, $B_{\theta b}^a (r,\theta)$, $C^a_{bc}$, we set
\begin{itemize}
\item for $n \neq 0$, $(\widetilde{\phi}^a_{\theta})_{n;0} = (\phi^{a(0)}_\theta)_{n}$, $(\widetilde{\phi}^a_{\theta})_{n;k} = 0 $ if $k > 0$;
\item $(\widetilde{\phi}_{r}^a)_{n;k}$ are to be ``dynamically'' found from \eqref{domegaFourier};
\item for $n=0$ $(\widetilde{\phi}^a_r)_{0;k} = 0$;
\item $(\widetilde{\phi}_\theta^a)_{0;k}$ are to be dynamically found from \eqref{domegaFourier}.
\end{itemize}
Graphically, this can be summarized the following way:\begin{center}\includegraphics[scale=0.97]{"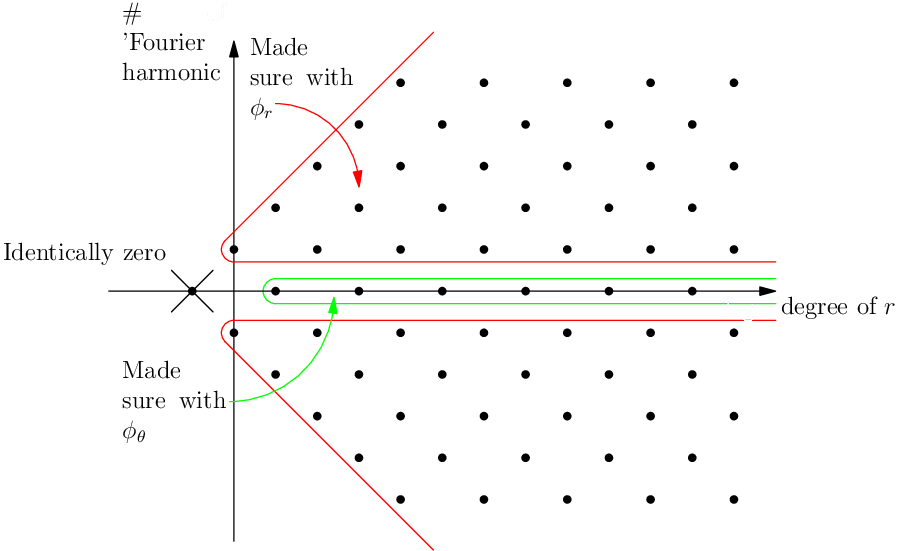"}\end{center}(Here, the bullets symbolize the $N$--dimensional vectors which are the $r^{|n| + 2k -1} e^{i n \theta}$ components of equation \eqref{domegaFourier}).

We see that with all the decisions made, the equations determine the $(\widetilde{\phi}_\theta^a)_{n;k}$ and $(\widetilde{\phi}_r^a)_{n;k}$ completely, except for $(\widetilde{\phi}_\theta^a)_{0;0}$. This last component is to be found from boundary conditions:
\[\label{phi00} (\widetilde{\phi}_\theta^a)_{0;0} = \phi_\theta^{a(0)} - (\widetilde{\phi}_\theta^{a})_{0;1} - (\widetilde{\phi}_\theta^{a})_{0;2} - \ldots = \widetilde{\phi}_\theta^{a(0)} - \sum_{k=1}^{\infty}(\widetilde{\phi}_\theta^{a})_{0;k}\]
(notice that we took care of all the boundary conditions when we were picking an ansatz for $(\widetilde{\phi}_r^a)_{n;k}$ except for $n=0$). We therefore run out of variables: there are no more free parameters to ensure $(\widetilde{\phi}_\theta^a)_{0;0} = 0$, which is necessary for smoothness! This can be seen as a manifestation of the above--described conjectural phenomenon: the solvability of extension problem is controlled by vanishing of a single $N$--component functional of boundary conditions. This last value of $(\widetilde{\phi}_\theta^a)_{0;0}$ presents therefore the obstruction to extension we were looking for, and can be called a 2--dimensional integral:
\[ P^{2}\exp{\int \omega}[\phi^{0}] = (\widetilde{\phi}_\theta^a)_{0;0} = \phi_\theta^{a(0)} - \sum_{k=1}^{\infty}(\widetilde{\phi}_\theta^{a})_{0;k},\]
where $(\phi_\theta^a)_{0;k}$ are to be found order--by--order from \eqref{domegaFourier}. Notice that as the equations for $(\widetilde{\phi}_\theta^{a})_{0;k}$ contain $(\widetilde{\phi}_\theta^a)_{0;0}$, the relation \eqref{phi00} is to be seen as an equation on $(\widetilde{\phi}_\theta^a)_{0;0}$ rather than a closed formula for it; because of its shape, however, it can be solved by reiteration over $(\widetilde{\phi}_\theta^a)_{0;0} $, which after expansion is identical with solution in terms of series in $\omega$ and $\phi^{(0)}$.

Perhaps a little demonstration of how this whole machinery works is in order. Consider the case of abelian 2--form $\omega$; it contains the components of $A_{r\theta} = r \sum (A_{r\theta})_{n;k} e^{i n \theta} r^{|n| + 2k +1}$ (the color index has been suppressed as it runs over only one value), while $B_{r\bullet}^{\bullet} = B_{\theta\bullet}^{\bullet} = C_{\bullet\bullet}^{\bullet} = 0$. According to the description of the extension problem, we have to fix a boundary 1--form $\phi^{(0)} = \left(\sum (\phi_\theta^{(0)})_n e^{i n \theta}\right) \mathrm{d} \theta$ defined at $r=1$. We then need to come up with two series of quantities $(\widetilde{\phi}_\theta)_{n;k}$ and $(\widetilde{\phi}_r)_{n;k}$ such that $\sum\limits_{k} (\widetilde{\phi}_\theta)_{n;k} = (\phi_\theta^{(0)})_{n}$ for every $n$ and
\[ \partial_r \widetilde{\phi}_\theta - \partial_\theta \widetilde{\phi}_r + A_{r\theta} = 0\]
which is equivalent to
\[ (|n| + 2 k) (\widetilde{\phi}_\theta)_{n;k} - i n (\widetilde{\phi}_r)_{n;k} + (A_{r\theta})_{n;k-1} = 0\]
for every $n$ and $k$. (Notice the shift of $k$ in the last term: while $(\widetilde{\phi}_\theta)_{n;k}$ and $(\widetilde{\phi}_r)_{n;k}$ contribute with $r^{|n|+2k-1}$, $(A_{r\theta})_{n;k}$ contributes with $r^{n+2k+1}$ because it comes in combination $A_{r\theta}\, r \mathrm{d}r \wedge \mathrm{d}\theta$).

For $n \neq 0$ we set, according to our ansatz, $(\widetilde{\phi}_\theta)_{n;k}$ to be $0$ if $k > 0$, and $(\widetilde{\phi}_\theta)_{n;k} = (\phi_\theta^{(0)})_n$. $(\widetilde{\phi}_r)_{n;k}$ therefore should be
\[ (\widetilde{\phi}_r)_{n;0} = - \frac{i}{n} |n|\phi_\theta^{0}\]

\[ (\widetilde{\phi}_r)_{n;0} = - \frac{i}{n} (A_{r\theta})_{n;k}\]
With the $n \neq 0$ conditions satisfied, we now go for the $n = 0$ ones. We have to set, according to our ansatz, $(\widetilde{\phi}_r)_{0;0}$ to be $0$. $(\widetilde{\phi}_\theta)_{0;k}$ therefore should be
\[ (\widetilde{\phi}_\theta)_{0;k} = - \frac{1}{2k} (A_{r\theta})_{0;k-1}\]

Therefore, the obstacle to extension is $(\widetilde{\phi}_\theta)_{0;0}$ set from initial conditions:
\[ (\widetilde{\phi}_\theta)_{0;0} = (\phi_\theta^{(0)})_{0} - \sum_{k=1}^{\infty} (\widetilde{\phi}_\theta)_{0;k} = (\widetilde{\phi}_\theta^{(0)})_{0} + \sum_{k=1}^{\infty} \frac{(A_{r\theta})_{0;k-1}}{2k}\]
Using the fact that $1/n = \int\limits_0^1 r^{n+1} \mathrm{d}r$, we can rewrite this expression as
\[\begin{aligned}
& P^{(2)}\exp{\int A}[\phi^{0}] = (\phi_\theta^{(0)})_{0} + \sum_{k=1}^{\infty} \frac{(A_{r\theta})_{0;k-1}}{2k} =\\
& (\phi_\theta^{(0)})_{0} + \sum_{k=0}^{\infty} \frac{(A_{r\theta})_{0;k}}{2k+2} =(\phi_\theta^{(0)})_{0} + \sum_{k=0}^{\infty} (A_{r\theta})_{0;k}) \int\limits_0^1 r^{2k+1} \mathrm{d}r=(\phi_\theta^{(0)})_{0} + \int\limits_{0}^{1} \left(\sum_{k=0}^{infty} (A_{r\theta})_{0;k} r^{2k}) \right) r\; \mathrm{d}r=\\
& (\phi_\theta^{(0)})_{0} + \int\limits_{0}^{1} (A_{r\theta})_0 \mathrm{d}r
\end{aligned}\]
Now, as by definition $(\phi_\theta^{(0)})_{0} = \frac{1}{2\pi} \oint \phi_\theta^{(0)} \mathrm{d}\theta$, $(A_{r\theta})_0(r) = \frac{1}{2\pi} \oint A_{r\theta}(r,\theta) \mathrm{d}\theta$,
\[\begin{aligned}
\label{AbConsCheck}&P^{(2)}\exp{\int A}[\phi^{0}] = \frac{1}{2\pi} \left( \oint \phi_\theta^{(0)} \mathrm{d}\theta + \int\limits_0^1 (\oint \phi_{r\theta} \mathrm{d}\theta)r \mathrm{d}r\right)=\\
& = \frac{1}{2 \pi} \left(\int\limits_{\partial D} \phi^{0} + \int\limits_{D} A \right)
\end{aligned}\]
We see that the usual two--dimensional integration formula is restored by our order--by--order calculations.

A relevant remark to be made here is the following. Notice that the result of our derivations, the formula \eqref{AbConsCheck} is much more invariant than our way to obtain it. Namely, if any disk reparametrization preserving the overall ``Moscow--like'' shape of radial coordinates is made, then both the Fourier components $(\widetilde{\phi} | \omega)_n$ and their radial Taylor components $(\widetilde{\phi} | \omega)_{n;k}$ are going to be mixed in a complicated fashion (just think about the horrible effect of substitutions $\theta \mapsto \theta + \alpha(r) \sin \theta$, $r \mapsto r + \beta(\cos \theta) r (1-r)$). The result, however, is insensitive to any coordinate changes inside the disk (provided that form components are changed appropriately). This is in parallel with the fact that the zeroth Fourier component of a $1$--form on the circle is much more invariant than all the others; analogous effects are to be expected in the general case as well.

\section{Assault on formula for obstacle 2: Taylor coefficient matching}\label{assault2}

In the previous section, the problem of finding the relations between boundary values of an $\omega$--flat splitting was approximated by the problem of finding the relations between its Fourier coefficients, starting from low--frequency ones and going to higher frequencies. This section is occupied with a sort of complementary approach: we study the non--abelian integration over an infinitesimal domain, which naturally leads to approximating the space of splittings with the space of their Taylor coefficients.

The story unfolds on the following triangular domain:\begin{center}  \includegraphics{"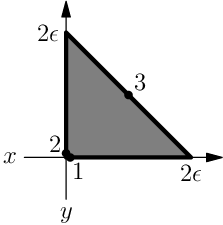"}\end{center}The approximation we are interested in is $\epsilon \rightarrow 0$. In order to exploit this limit, we expand $\widetilde{\phi}_{x|y}^a$\footnote{$x|y$ stands for ``$x$ or $y$'', just like $\pm$ stands for ``$+$ or $-$''} in $x$ and $y$ around the point $(0,0)$:
\[\widetilde{\phi}_{x|y}^a(x,y) = \sum\limits_{n=0}^{\infty} \frac{1}{m! n!} x^m y^n \left(\partial_x^m \partial_y^n \widetilde{\phi}_{x|y}^a\right)(0,0)\]
In what follows we are going to denote $\left(\partial_x^m \partial_y^n \widetilde{\phi}_{x|y}^a\right)(0,0) = \partial_x^m \partial_y^n \widetilde{\phi}_{x|y}^a$ without an explicit reference to point $(0,0)$. The same convention applies whenever $\partial_{x|y}$ acts on an expression: $(\partial_x + \partial_y)\partial_x X$ is a shorthand for $\left(\partial^2_x X + \partial_y \partial_x X\right) (0,0)$.

In these ``coordinates'' the boundary data consists of three pieces --- the restrictions of $\widetilde{\phi}$ onto the sides of the triangle. This means we are given the function $\widetilde{\phi}^a_x$ at the points $(0 \leqslant x \leqslant 2 \epsilon; y = 0)$ only, the function $\widetilde{\phi}^a_y$ at the points $(x = 0; 0 \leqslant y \leqslant 2 \epsilon)$ only, and their difference $\widetilde{\phi}^a_x - \widetilde{\phi}^a_y$ at the points $(\epsilon+t; \epsilon -t)$, $-\epsilon \leqslant t \leqslant \epsilon$ only. These pieces of boundary data will be denoted $\phi^a_1(x)$, $\phi^a_2(y)$, and $\phi^a_3(t)$, in that order. We expand these single--variable functions around the points $1$, $2$, and $3$ with coordinates $(0,0)$, $(0,0)$, and $(\epsilon,\epsilon)$, respectively:
\[ \phi_1^a(x) = \sum\limits_{k=0}^{\infty} \frac{1}{k!} x^k \partial_1^{k} \phi_1^a \]

\[ \phi_2^a(y) = \sum\limits_{k=0}^{\infty} \frac{1}{k!} y^k \partial_2^{k} \phi_2^a \]

\[ \phi_3^a(t) = \sum\limits_{k=0}^{\infty} \frac{1}{k!} t^k \partial_3^k \phi_3^a \]
They can be expressed through $(0,0)$--rooted Taylor coefficients of $\widetilde{\phi}^a_{x|y}$ as series in $\epsilon$:
\[ \partial_1^{k} \phi_1^a = \partial_x^{k} \widetilde{\phi}_x^a  \]

\[ \partial_2^{k} \phi_2^a = \partial_y^{k} \widetilde{\phi}_y^a \]

\[ \partial_3^k \phi_3^a = \sum_{n = 0}^{\infty} \epsilon^n (\partial_x + \partial_y)^n (\partial_x - \partial_y)^k (\widetilde{\phi}_x^a - \widetilde{\phi}_y^a)\]
In particular, for $k=0$ $\widetilde{\phi}^a_x = \phi^a_1$ and $\widetilde{\phi}^a_y = \phi^a_2$ --- but beware that these symbols denote the values of $\widetilde{\phi}_{x|y}^a$/$\phi^a_{1|2}$ at $(0,0)$, and the equality is not valid under the differentiation sign! The logic of this convention is that the symbols $\partial_1^k \phi^a_1$, $\partial_2^k \phi^b_2$, and $\partial_3^k \phi^c_3$ are pieces of boundary data, while $\partial_x^m \partial_y^n \widetilde{\phi}^a_x$ and $\partial_x^m \partial_y^n \widetilde{\phi}^b_y$ are bulk quantities: they determine the behavior of $\widetilde{\phi}$ at the interior points of the triangle. Be careful with these notations: while $\partial_1^2 \phi^a_1 = \partial_x^2 \widetilde{\phi}^a_x$, the chunk $\partial_1 \partial_2 \phi^a_1$ is meaningless as $\phi^a_1$, being the restriction of $\widetilde{\phi}$ onto the interval $(0,0)-(2\epsilon,0)$, is not defined at $y \neq 0$ and therefore cannot be differentiated over $y$! Notice also that for a large order $k$ the boundary data contain only $\sim 3 k$ coefficients, while $\sim k^2/2$ coefficients are necessary to describe $\widetilde{\phi}_{\mu}^a$ completely to the order $k$.

According to the main narrative, the differential equation
\[\label{domegaa2} \partial_x \widetilde{\phi}_y^a - \partial_y \widetilde{\phi}_x^a + A_{xy}^a + B_{xb}^a \widetilde{\phi}_y^b - B_{yb}^a \widetilde{\phi}_x^b + C^a_{bc} \widetilde{\phi}_x^b \widetilde{\phi}_y^c = 0 \]
should induce a single (multicomponent) relation on boundary data:
\[\label{inttaylorfull} I_{(A,B,C)}^a[\phi_1^b,\partial_1 \phi_1^{b_1}, \partial_1^2 \phi_1^{b_2}, \ldots, \phi_2^c, \partial_2 \phi_2^{c_1}, \ldots, \phi_3^d, \partial_3 \phi_3^{d_1}, \ldots] = 0\]
Suppose, because of the infinitesimalitude of the triangle, we have truncated our boundary data to the order--$k$ in all the derivatives; what should we expect as a partial result in the spirit of \eqref{inttaylorfull}? Of course it would be naive to look for an exact relation between the boundary derivatives at different points, but what is the best precision we can achieve? It turns out that the most reasonable expectation is:\begin{itemize}\item There exists a quantity $I_n^d[\phi_1^a,\ldots,\partial_1^{n-1}\phi_1^{a_{n-1}}, \phi_2^b, \ldots, \partial_2^{n-1} \phi_2^{b_{n-1}}, \phi_3^{c}, \ldots, \partial_3^{n-1} \phi_3^{c_{n-1}}]$ that is not a priori divisible by $\epsilon$, depends on no derivative higher than $\partial^{n-1}_{1|2|3}$ on each boundary segment of $1$, $2$, $3$, and which is of order $\sim \epsilon^{n}$ for $(A,B,C)$--flat splittings (i.e., given \eqref{domegaa2}).\end{itemize}In particular, this is exactly what happens if we approximate all the boundary forms by polynomials in the \emph{usual} Stokes formula:
\[\begin{aligned}\frac{1}{\epsilon} I_{A}[\phi] = \frac{1}{\epsilon} \int\limits_{-\epsilon}^{\epsilon} \phi_3(t) dt - \frac{1}{\epsilon} \int\limits_{0}^{2\epsilon} \phi_1(x) dx + \frac{1}{\epsilon} \int\limits_{0}^{2\epsilon} \phi_2(y) d y + \frac{1}{\epsilon}\int\limits_{\bigtriangleup} A_{xy} dx \wedge dy \approx\\
 \approx\frac{1}{\epsilon} \sum\limits_{i=0}^{n} \frac{1}{i!} \partial_3^i \phi_3 \frac{\epsilon^{i+1} - (- \epsilon)^{i+1}}{i+1} - \frac{1}{\epsilon} \sum\limits_{j=0}^{n} \frac{1}{j!} \partial_1^j \phi_1 \frac{(2 \epsilon)^{j+1}}{j+1} + \frac{1}{\epsilon} \sum\limits_{k=0}^{n}\frac{1}{k!} \partial_1^k \phi_1 \frac{(2 \epsilon)^{k+1}}{k+1} + \frac{1}{\epsilon} \int\limits_{\bigtriangleup} A_{xy} dx \wedge dy\end{aligned}\]
Here the sums approximate the quantity $\oint \phi = - \int\limits_{\bigtriangleup} A$ up to the order $\epsilon^{n+2}$, therefore, the last expression is of order $\epsilon^{n+1}$. The coefficient $\frac{1}{\epsilon}$ was introduced so that the expansion starts with $\phi_3 - \phi_1 + \phi_2 + \ldots$ as opposed to $\epsilon \phi_3 - \epsilon \phi_1 + \epsilon \phi_2 + \ldots$.

What follows is a demonstration of the same phenomenon in the non--abelian case. More specifically, we construct the appropriate quantities $I_k$ for $k=1,2,3$ and show how to construct $I_4$ without writing it down due to space constraints.

The construction process starts with the following simple fact: any splitting $\phi$, whether $\omega$--flat or not, satisfies the approximate equality
\[ I_1^a[\phi] = \phi_3^a - \phi_1^a + \phi_2^a \sim \epsilon \]
because $\phi_3^a = (\widetilde{\phi}^a_x - \widetilde{\phi}^a_y) (\epsilon,\epsilon) = \phi^a_1 - \phi^a_2 + \epsilon (\ldots)$. If we expand this quantity up to the order $\epsilon$, it becomes
\[ (\widetilde{\phi}^a_x - \widetilde{\phi}^a_y) - \widetilde{\phi}^a_x + \widetilde{\phi}^a_y + \epsilon (\partial_x \widetilde{\phi}^a_x + \partial_y \widetilde{\phi}^a_x - \partial_x \widetilde{\phi}^a_y -\partial_y \widetilde{\phi}^a_y) + \epsilon^2(\ldots) \]
Therefore, for unconstrained forms $\widetilde{\phi}$ this approximate equality cannot be promoted to one of order $\sim \epsilon^2$, as the residual term is not fixed by boundary data (the combinations $\partial_1 \phi^a_1 = \partial_x \widetilde{\phi}^a_x$, $\partial_2 \phi^a_2 = \partial_y \widetilde{\phi}^a_y$, $\partial_3 \phi^a_3 = \partial_x \widetilde{\phi}^a_x - \partial_y \widetilde{\phi}^a_y - \partial_x \widetilde{\phi}^a_y - \partial_y \widetilde{\phi}^a_y$ and $\partial_x \widetilde{\phi}^a_x + \partial_y \widetilde{\phi}^a_x - \partial_x \widetilde{\phi}^a_y -\partial_y \widetilde{\phi}^a_y$ which we have obtained are literally linearly independent). However, given the principal equation \eqref{domegaa2}, such a promotion can be done:
\[ \partial_x \widetilde{\phi}_y^a - \partial_y \widetilde{\phi}_x^a + A^a + B_{xb}^a \widetilde{\phi}_y^b - B_{yb}^a \widetilde{\phi}_x^b + C^a_{bc} \widetilde{\phi}_x^b \widetilde{\phi}_y^c = 0 \]
(here $A^a_{xy}$ was shortened to $A^a$), therefore, taking this equation at $(0,0)$
\[ \partial_x \widetilde{\phi}^a_x + \partial_y \widetilde{\phi}^a_x - \partial_x \widetilde{\phi}^a_y -\partial_y \widetilde{\phi}^a_y = \partial_1 \phi^a_1 - \partial_2\phi^a_2 + (A^a + B_{xb}^a \phi^b_2 - B_{yb}^a \phi^b_1 + C^a_{bc} \phi^b_1 \phi^c_2)\]
which implies the expression
\[\label{I2} I_2^a[\phi] = \phi_3^a - \phi_1^a + \phi_2^a - \epsilon \partial_1 \phi_1^a + \epsilon \partial_2 \phi_2^a - \epsilon (A^a + B_{xb}^a \phi^b_2 - B_{yb}^a \phi^b_1 + C^a_{bc} \phi^b_1 \phi^c_2 ) \sim \epsilon^2 \]
is of the order $\epsilon^2$.

To move on to the next order, we follow the same line of thought: expand the $I_2[\phi]$ function we have already found to the next power of $\epsilon$ in $\widetilde{\phi}$ variables and try to express the residual term through boundary data. The expansion of $I_2$ to the order $\epsilon^2$ has the form
\[\begin{aligned}I_2[\phi] = \widetilde{\phi}^a_x - \widetilde{\phi}^a_y + \epsilon \partial_x \widetilde{\phi}^a_x + \epsilon(-\partial_y \widetilde{\phi}^a_x + \partial_x \widetilde{\phi}^a_y) - \epsilon \partial_y \widetilde{\phi}^a_y-\\
- \widetilde{\phi}^a_x + \widetilde{\phi}^a_y - \epsilon \partial_x \widetilde{\phi}^a_x + \epsilon \partial_y \widetilde{\phi}^a_y + \epsilon(A^a + B^a_{xb} \widetilde{\phi}^b_y - B^a_{yb} \widetilde{\phi}^b_x + C^a_{bc} \widetilde{\phi}^b_x \widetilde{\phi}^c_y)+\\
+ \frac{\epsilon^2}{2} (\partial_x^2 \widetilde{\phi}^a_x + 2 \partial_x \partial_y \widetilde{\phi}^a_x + \partial_y^2\widetilde{\phi}^a_x - \partial_x^2 \widetilde{\phi}^a_y - 2 \partial_x \partial_y \widetilde{\phi}^a_y - \partial_y^2 \widetilde{\phi}^a_y) + \epsilon^3(\ldots)\end{aligned}\]
The first two lines cancel each other as expected from the considerations above. To deal with the last one, notice the following. If we add to $I_2$ the term $\frac{\epsilon^2}{6} \partial_3^2 \phi^a_3$ (which is a part of boundary data), the $\epsilon^0$, $\epsilon^1$ cancellations are not affected, but the $\epsilon^2$ residual term becomes
\[\begin{aligned} \frac{\epsilon^2}{6} (4 \partial_x^2 \widetilde{\phi}^a_x + 4 \partial_x \partial_y \widetilde{\phi}^a_x + 4 \partial_y^2 \widetilde{\phi}^a_x - 4 \partial_x^2 \widetilde{\phi}^a_y - 4 \partial_x \partial_y \widetilde{\phi}^a_y - 4 \partial_y^2 \widetilde{\phi}^a_y ) =\\
= \frac{2}{3} \epsilon^2 \partial_1^2 \phi^a_1 - \frac{2}{3} \epsilon^2 \partial_2^2 \phi^a_2 + \frac{2}{3} \epsilon^2 \partial_x (\partial_y \widetilde{\phi}^a_x - \partial_x \widetilde{\phi}^a_y) + \frac{2}{3} \epsilon^2 \partial_y (\partial_y \widetilde{\phi}^a_x - \partial_x \widetilde{\phi}^a_y)\end{aligned}\]
As the equation \eqref{domegaa2} is supposed to hold identically on the domain of integration, we can use it under the sign of partial derivative:
\[ \label{res2}\begin{aligned}  \mathrm{residual}_2 = \frac{2}{3} \epsilon^2 \partial_1^2 \phi^a_1 - \frac{2}{3} \epsilon^2 \partial_2^2 \phi^a_2 +\\
  + \frac{2}{3} \epsilon^2 \partial_x (A^a + B^a_{xb} \widetilde{\phi}^b_y - B^a_{yb} \widetilde{\phi}^b_x + C^a_{bc} \widetilde{\phi}^b_x \widetilde{\phi}^c_y) + \frac{2}{3} \epsilon^2 \partial_y (A^a + B^a_{xb} \widetilde{\phi}^b_y - B^a_{yb} \widetilde{\phi}^b_x + C^a_{bc} \widetilde{\phi}^b_x \widetilde{\phi}^c_y)=\\
  = \frac{2}{3} \epsilon^2 \partial_1^2 \phi^a_1 - \frac{2}{3} \epsilon^2 \partial_2^2 \phi^a_2 + \frac{2}{3} \epsilon^2 (\partial_x A^a - \partial_y A^a) + \\
  +\frac{2}{3} \epsilon^2 (\partial_x B^a_{xb} + \partial_y B^a_{xb} ) \phi^b_2 - \frac{2}{3} \epsilon^2 (\partial_x B^a_{yb} + \partial_y B^a_{yb}) \phi^b_1 +  \frac{2}{3} \epsilon^2 \left(B^a_{xb} (\partial_2 \phi_2)^b - B^a_{yb} (\partial_1 \phi_1)^b\right) +\\
  +\frac{2}{3}\epsilon^2 (\partial_x C^a_{bc} + \partial_y C^a_{bc}) \phi_1^b \phi_2^c + \frac{2}{3} \epsilon^2 C^a_{bc} ((\partial_1 \phi_1)^b \phi_2^c + \phi_1^b (\partial_2 \phi_2)^c)+\\
  +\frac{2}{3} \epsilon^2 B^a_{xb} (\partial_x \widetilde{\phi}_y)^b - \frac{2}{3} \epsilon^2 B^a_{yb} (\partial_y \widetilde{\phi}_x)^b + \frac{2}{3} \epsilon^2 C^a_{bc} (\partial_y \widetilde{\phi}_x)^b \widetilde{\phi}_y^c + \frac{2}{3} \epsilon^2 C^a_{bc} \widetilde{\phi}_x^b (\partial_x \widetilde{\phi}_y)^c\end{aligned}\]
We therefore observe a phenomenon absent in the abelian case: as $\partial_x \widetilde{\phi}^a_y - \partial_y \widetilde{\phi}^a_x$ now depends on $\widetilde{\phi}$, the terms obtained from the derivatives of that are no longer obliged to be boundary ones. This is seen from the last line: the quantities $\partial_y \widetilde{\phi}_x^a$ and $\partial_x \widetilde{\phi}_y^a$ \emph{separately} are not part of boundary data $\phi_{1,2,3}$. Using the condition \eqref{domegaa2}, we can, however, express them through boundary data. Indeed,
\[\left\{\begin{aligned}& \partial_x \widetilde{\phi}^a_x - \partial_y \widetilde{\phi}^a_x - \partial_x \widetilde{\phi}^a_y + \partial_y \widetilde{\phi}^a_y = \partial_3 \phi^a_3 \\
& \partial_y \widetilde{\phi}^a_x - \partial_x \widetilde{\phi}^a_y = A^a + B^a_{xb} \phi^b_2 - B^a_{yb} \phi^b_1 + C^a_{bc} \phi^b_1 \phi^c_2\end{aligned}\right.\]
therefore
\[\label{dxphiy}\left\{\begin{aligned}& \partial_y \widetilde{\phi}^a_x = - \frac{1}{2} \partial_3 \phi^a_3 + \frac{1}{2} \partial_1 \phi^a_1 + \frac{1}{2} \partial_2 \phi^a_2 + \frac{1}{2} \left(A^a + B^A_{xb} \phi^b_2 - B^a_{yb} \phi^b_1 + C^a_{bc} \phi^b_1 \phi^c_2\right)\\
& \partial_x \widetilde{\phi}^a_y = - \frac{1}{2} \partial_3 \phi^a_3 + \frac{1}{2} \partial_1 \phi^a_1 + \frac{1}{2} \partial_2 \phi^a_2 - \frac{1}{2} \left(A^a + B^a_{xb} \phi^b_2 - B^a_{yb} \phi^b_1 + C^a_{bc} \phi^b_1 \phi^c_2\right)\end{aligned}\right.\]
By inserting this into the expression \eqref{res2}, we obtain the quantity
\[\begin{aligned} I_3^a[\phi] = \phi^a_3 - \phi^a_1 + \phi^a_2 - \epsilon \partial_1 \phi^a_1 + \epsilon \partial_2 \phi^a_2 - \epsilon (A^a + B^a_{xb} \phi^b_2 - B^a_{yb} \phi^b_1 + C^a_{bc} \phi_1^b \phi_2^c )+\\
+ \frac{1}{6} \epsilon^2 \partial_3^2 \phi^a_3 -\frac{2}{3}\epsilon^2 \partial_1^2 \phi^a_1 + \frac{2}{3} \epsilon^2 \partial_2^2 \phi^a_2 - \frac{2}{3} \epsilon^2 (\partial_x A^a + \partial_y A^a)-\\
- \frac{2}{3} \epsilon^2 (\partial_x B^a_{yb} + \partial_y B^a_{yb}) \phi^b_1 + \frac{2}{3} \epsilon^2 (\partial_x B^a_{xb} + \partial_y B^a_{xb}) \phi^b_2 - \frac{1}{3} \epsilon^2 (B^a_{xb} (3 \partial_2 \phi^b_2 + \partial_1 \phi^b_1) - B^a_{yb} (3 \partial_1 \phi^b_1 + \partial_2 \phi^b_2))-\\
- \frac{2}{3} \epsilon^2 (\partial_x C^a_{bc} + \partial_y C^a_{bc}) \phi^b_1 \phi^c_2 - \frac{2}{3} \epsilon^2 C^a_{bc} \left((3 \partial_1 \phi^b_1 + \partial_2 \phi^b_2) \phi^c_2 + \phi^b_1 (3 \partial_2 \phi^c_2 + \partial_1 \phi^c_1)\right)+\\
+ \frac{1}{3} \epsilon^2 (B^a_{xc} + C^a_{bc} \phi^b_1) \left(\partial_3 \phi^c_3 + (A^c + B^c_{xd} \phi^d_2 - B^c_{yd} \phi^d_1 + C^c_{de} \phi^d_1 \phi^e_2)\right)-\\
- \frac{1}{3} \epsilon^2 (B^a_{yc} + C^a_{bc} \phi^b_2) \left(\partial_3 \phi^c_3 - (A^c + B^c_{xd} \phi^d_2 - B^c_{yd} \phi^d_1 + C^c_{de} \phi^d_1 \phi^e_2)\right)\sim \epsilon^3\end{aligned}\]
which is of order $\sim \epsilon^3$ for $(A,B,C)$--flat splittings $\phi^a_\mu$. (In the last line we have used the identity $C^a_{bc} X^b \phi_2^c = - C^a_{bc} \phi_2^b X^c$ to shorten up the expression.)

As is evident, the complete expression for the integral gets quite complicated already for the quadratic approximation. In order not to overwhelm the exposition, we step aside from deriving a complete expression in the next order and are going to merely show that such an expression exists. To do this, it suffices to demonstrate that the $\epsilon^3$ residual term of $I_3$ is expressable through boundary values on \eqref{domegaa2}. We start the demonstration by writing down only the $\phi_3$--dependent part of $I_3$, as this part is the only one that contributes to further expansion:
\[ I_3^a[\phi] = \phi^a_3 + \frac{1}{6} \epsilon^2 \partial_3 \phi^a_3 + \frac{1}{3} \epsilon^2 (B^a_{xc} + C^a_{bc} \phi^b_1 - B^a_{yc} - C^a_{bc} \phi^b_2) \partial_3 \phi^c_3 + f^a(\phi^{b}_1,\partial_1 \phi^{b_1}_1, \partial_1^2 \phi^{b_2}_1, \phi^{c}_2,\partial_2 \phi^{c_1}_2, \partial_2^2 \phi^{c_2}_2)\]
From this, it is possible to read off the cubic order residual term (the two previous ones cancel out):
\[\begin{aligned}I_3^a[\phi] = \frac{\epsilon^3}{6} (\partial^3_x \widetilde{\phi}^a_x + 3 \partial_x^2 \partial_y \widetilde{\phi}^a_x+ 3 \partial_x \partial_y^2 \widetilde{\phi}^a_x + \partial_y^3 \widetilde{\phi}^a_x - \partial^3_x \widetilde{\phi}^a_y - 3 \partial_x^2 \partial_y \widetilde{\phi}^a_y- 3 \partial_x \partial_y^2 \widetilde{\phi}^a_y - \partial_y^3 \widetilde{\phi}^a_y) +\\
+\frac{\epsilon^3}{6} (\partial_x^3 \widetilde{\phi}^a_x - \partial_x^2 \partial_y \widetilde{\phi}^a_x - \partial_x \partial_y^2 \widetilde{\phi}^a_x + \partial_y^3 \widetilde{\phi}^a_x - \partial_x^3 \widetilde{\phi}^a_x + \partial_x^2 \partial_y \widetilde{\phi}^a_y + \partial_x \partial_y^2 \widetilde{\phi}^a_y - \partial_y^3 \widetilde{\phi}^a_y) +\\
+\frac{\epsilon^3}{3} (B^a_{xc} + C^a_{bc} \phi^b_1 - B^a_{yc} - C^a_{bc} \phi^b_2) (\partial_x^2 \widetilde{\phi}^c_x - \partial_y^2 \widetilde{\phi}^c_x - \partial_x^2 \widetilde{\phi}^c_y +\partial_y^2 \widetilde{\phi}^c_y) + \epsilon^4(\ldots)=\\
= \frac{\epsilon^3}{6} (2 \partial_x^3 \widetilde{\phi}^a_x + 2 \partial_x^2 \partial_y \widetilde{\phi}^a_x - 2 \partial_x^3 \widetilde{\phi}^a_y + 2 \partial_x \partial_y^2 \widetilde{\phi}^a_x - 2 \partial_x^2 \partial_y \widetilde{\phi}^a_y + 2 \partial_y^3 \widetilde{\phi}^a_x - 2 \partial_x \partial_y^2 \widetilde{\phi}^a_y - 2 \partial_y^3 \widetilde{\phi}^a_y) + \\
+\frac{\epsilon^3}{3} (B^a_{xc} + C^a_{bc} \phi^b_1 - B^a_{yb} - C^a_{bc} \phi^b_2) (\partial_x^2 \widetilde{\phi}^c_x - \partial_y^2 \widetilde{\phi}^c_x - \partial_x^2 \widetilde{\phi}^c_y +\partial_y^2 \widetilde{\phi}^c_y) + \epsilon^4(\ldots) =\\
=\frac{\epsilon^3}{6} \left[2 \partial_1^3 \phi^a_1 + 2 \partial_x^2 (A^a + B^a_{xb} \widetilde{\phi}^b_y - B^a_{yb} \widetilde{\phi}^b_x + C^a_{bc} \widetilde{\phi}^b_x \widetilde{\phi}^c_y) + 2 \partial_x \partial_y  (A^a + B^a_{xb} \widetilde{\phi}^b_y - B^a_{yb} \widetilde{\phi}^b_x + C^a_{bc} \widetilde{\phi}^b_x \widetilde{\phi}^c_y)  +\right. \\
+ \left. 2 \partial_y^2 (A^a + B^a_{xb} \widetilde{\phi}^b_y - B^a_{yb} \widetilde{\phi}^b_x + C^a_{bc} \widetilde{\phi}^b_x \widetilde{\phi}^c_y) - 2 \partial_2^3 \widetilde{\phi}_2^a\right] +\\
+\frac{\epsilon^3}{3} (B^a_{xc} + C^a_{bc} \phi^b_1 - B^a_{yc} - C^a_{bc} \phi^b_2) (\partial_x^2 \widetilde{\phi}^c_x - \partial_y^2 \widetilde{\phi}^c_x - \partial_x^2 \widetilde{\phi}^c_y +\partial_y^2 \widetilde{\phi}^c_y) + \epsilon^4(\ldots)\end{aligned}\]
Thanks to formulas \eqref{dxphiy}, we know that all the subexpressions depending on the first derivatives of $\widetilde{\phi}$ and not on higher ones can be reduced to boundary data. Therefore the most troublesome terms are the ones dependent on $\partial_y^2 \widetilde{\phi}^a_x$, $\partial_x \partial_y \widetilde{\phi}^a_x$, $\partial_x^2 \widetilde{\phi}^a_y$, and $\partial_x \partial_y \widetilde{\phi}^a_y$ (as the terms $\partial_x^2 \widetilde{\phi}^a_x = \partial_1^2 \phi_1$ and $\partial_y^2 \widetilde{\phi}^a_y = \partial_2^2 \phi_2$ are already boundary). This means the $\epsilon^3$ residual term takes the form
\[\label{I4magic}\begin{aligned}  \frac{\epsilon^3}{3} (B^a_{xc} + C^a_{bc} \widetilde{\phi}^b_x) (\partial_x^2 \widetilde{\phi}^c_y + \partial_x \partial_y \widetilde{\phi}^c_y) - \frac{\epsilon^3}{3} (B^a_{yc} + C^a_{bc} \widetilde{\phi}^b_y) (\partial_y^2 \widetilde{\phi}^c_x + \partial_x\partial_y \widetilde{\phi}^c_x)-\\
  -\frac{\epsilon^3}{3} (B^a_{xc} + C^a_{bc} \phi^b_1 - B^a_{yc} - C^a_{bc} \phi^b_2) (\partial_y^2 \widetilde{\phi}^c_x + \partial_x^2 \widetilde{\phi}^c_y) + \\
  + \epsilon^3 f^a(\phi_1^{b},\partial_1 \phi_1^{b_1}, \partial_1^2 \phi_1^{b_2}, \partial_1^3 \phi_1^{b_3}, \phi_2^{c}, \partial_2 \phi_2^{c_1}, \partial_2^2 \phi_2^{c_2}, \partial_2^3 \phi_2^{c_3}, \phi_3^{d}, \partial_3 \phi_3^{d_1}) =\\
  =\frac{\epsilon^3}{3} (B^a_{xc} + C^a_{bc} \phi^b_1)(\partial_x \partial_y \widetilde{\phi}^c_y - \partial_y^2 \widetilde{\phi}^c_x) - \frac{\epsilon^3}{3} (B^a_{yc} + C^a_{bc} \phi^b_2) (\partial_x \partial_y \widetilde{\phi}^c_x - \partial_x^2 \widetilde{\phi}^c_y) +\\
  +\epsilon^3 f^a(\phi_1^{b},\partial_1 \phi_1^{b_1}, \partial_1^2 \phi_1^{b_2}, \partial_1^3 \phi_1^{b_3}, \phi_2^{c}, \partial_2 \phi_2^{c_1}, \partial_2^2 \phi_2^{c_2}, \partial_2^3 \phi_2^{c_3}, \phi_3^{d}, \partial_3 \phi_3^{d_1}) = \\
  = -\frac{\epsilon^3}{3} (B^a_{xc} + C^a_{bc} \phi^b_1) \partial_y (A^c + B^c_{xd} \widetilde{\phi}^d_y - B^c_{yd} \widetilde{\phi}^d_x + C^c_{de} \widetilde{\phi}^d_x \widetilde{\phi}^e_y) -\\
  -\frac{\epsilon^3}{3} (B^a_{yc} + C^a_{bc} \phi^b_2) \partial_x (A^c + B_{xd}^c\widetilde{\phi}^d_y - B^c_{yd} \widetilde{\phi}^d_x + C^c_{de} \widetilde{\phi}^d_x \widetilde{\phi}^e_y)+\\
+ \epsilon^3 f^a(\phi_1^{b},\partial_1 \phi_1^{b_1}, \partial_1^2 \phi_1^{b_2}, \partial_1^3 \phi_1^{b_3}, \phi_2^{c}, \partial_2 \phi_2^{c_1}, \partial_2^2 \phi_2^{c_2}, \partial_2^3 \phi_2^{c_3}, \phi_3^{d}, \partial_3 \phi_3^{d_1})\end{aligned}\]
which contains only the first derivatives of $\widetilde{\phi}$ and is therefore packable into boundary--dependent shape. Therefore it is possible, in principle, to write down a function $I_4^a(\phi^b_1,\partial_1 \phi^{b_1}_1, \partial_1^2 \phi^{b_2}_1, \partial_1^3 \phi^{b_3}_1, \phi_2^c,\ldots,\partial_2^3 \phi^{c_3}_2, \phi^d_3, \ldots, \partial_3^3 \phi^{d_3}_3)$ which is $\sim \epsilon^4$ for $\widetilde{\phi}$ obeying \eqref{domegaa2}. Notice that in this order the boundary values of $\widetilde{\phi}$ do not fix it completely up to $\epsilon^3$ even given the \eqref{domegaa2}, but it is nevertheless possible to find a relation between these values.

For the fourth order, the phenomenon analogous to \eqref{I4magic} occurs for the terms depending on $\partial^3 \widetilde{\phi}$ Taylor coefficients; forcing it to happen subsequently for $\partial^2 \widetilde{\phi}$--dependent terms seems, however, to require an exploitation of ambiguity we cannot readily point out to. The relations \eqref{I2}--\eqref{I4magic} nevertheless seem to us to be of interest already.

An inquisitive reader might ask, why did not we proceed from the formula \eqref{I2} to larger triangles by the usual way of their dissection into smaller ones? The relation \eqref{I2} allows one, given the values of $\phi$ on two sides of an infinitesimal triangle, to find, within an error of order $\epsilon^2$, the value of $\phi$ on the third one, so when several triangles are glued by a common border segment, it should be possible, at least theoretically, to compose these expressions and get an integral over their union, etc.\begin{center}  \includegraphics{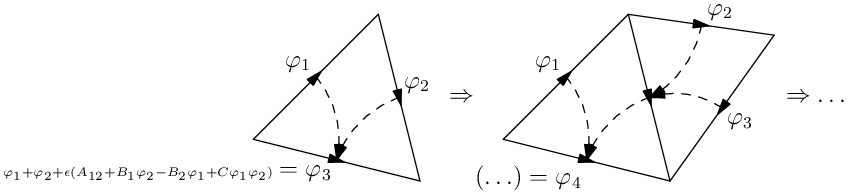}\end{center}However, when the triangulation becomes large enough to contain interior points, the following problem emerges. During the process of ``shrinking the triangles one--by--one'' the situation when only one side of the triangle carries a known $\phi$ is inevitable. As we have only one relation per plaquette, the values of $\phi$ on two unknown sides cannot be found from a known one; actually, we can voluntaristically assign the value of $\phi$ to one of the unprocessed sides and deduce the other one.\begin{center}  \includegraphics{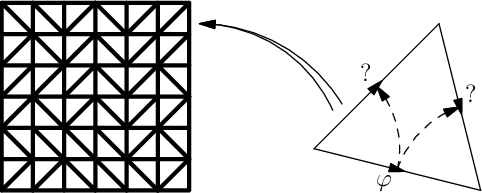}\end{center}This is a manifestation of a before--mentioned ambiguity: previously, it was discussed as the need for $\phi^a_y$ for parallel transport in \ref{sectransport}, and as the requirement to quess $\widetilde{\phi}^a_r$ in \ref{assault1}. For the triangulation approach, no reasonable principle that could guide the balance between two unknown sides is immediately available. Moreover, when the number of plaquettes is finite, we cannot even detect a singularity--like growth of $\phi$ which used to be our clue in \ref{assault1}, because the divergence only occurs when the number of resubstitutions tends to infinity. This difficulty obstructs the triangulation path of exponentiating the infinitesimal formula \eqref{I2} to some full--form one.

\section{The continuous Baker--Campbell--Hausdorff formula}\label{bch}

For our second classical example, flat connections, the order--by--order approach in \ref{assault1} leads to equations that get out of control long before any reasonable structure can be spotted. Nevertheless, the formula of the required type exists\cite{Magnus:1954zz}, and is, in fact, a continuous analog of BCH formula. We briefly recollect the formula together with its derivation; said derivation, however, relies heavily on the possibility of introducing additional (matrix) structure on top of Lie algebra--like one for classical connections, and so cannot be immediately generalized in the 2--connection direction. The formula in question is:
\[\label{CBCH}\begin{aligned}
 & \Phi(t) = \log P\exp \int\limits_0^t \phi(\tau) d\tau = \int\limits_0^t \frac{[\Phi(\tau),\cdot]}{e^{[\Phi(\tau),\cdot]} -1} \phi(\tau) d\tau = \\
 & = \int_0^t \phi(\tau) d\tau - \frac{1}{2}\int\limits_{0 < t_1 < t_2 < t} [\phi(t_2), \phi(t_1)] dt_1 dt_2 + \frac{1}{6} \int\limits_{0 < t_1 < t_2 < t_3 <t} [\phi(t_3),[\phi(t_2),\phi(t_1)]] d t_1 dt_2 d t_3 + \\
 & + \frac{1}{6} \int\limits_{0 < t_1 < t_2 < t_3 <t} [[\phi(t_3),\phi(t_2)],\phi(t_1)]] d t_1 dt_2 d t_3 + \ldots
\end{aligned}\]
 The usage of this formula goes as follows. The expression inside the integral $\frac{[\Phi(\tau),\cdot]}{e^{[\Phi(\tau),\cdot]}-1}$ is a matrix function in operator $[\Phi(\tau),\cdot]$ which is to be thought to be a mapping from matrix space to itself. It is to be understood as a series expansion $ \left(\frac{x}{e^{x}-1}\right)([\Phi(t),\cdot]) = (1 - \frac{x}{2} + \frac{x^2}{12} - \ldots)([\Phi(\tau),\cdot]) = 1 - \frac{1}{2} [\Phi(\tau),\cdot] + \frac{1}{12} [\Phi(\tau),[\Phi(\tau),\cdot]] + \ldots$ (so powers of $x$ are replaced by nested commutators with $\Phi$ of the same depth). Those composed commutators then act on $\phi(\tau)$ and are integrated. Therefore, the expression takes the form 
\[\begin{aligned}
\label{CBCHexpanded} \Phi(t) = \int\limits_0^t (1 - \frac{1}{2} [\Phi(\tau),\cdot] + \frac{1}{12} [\Phi(\tau),[\Phi(\tau),\cdot]] + \ldots) \phi(\tau) d\tau =\\
 =\int\limits_0^t (\phi(\tau) - \frac{1}{2} [\Phi(\tau),\phi(\tau)] + \frac{1}{12} [\Phi(\tau),[\Phi(\tau),\phi(\tau)]] + \ldots) d\tau =\\
 = \int \limits_0^t \phi(\tau) d\tau - \frac{1}{2} \int \limits_0^t [\Phi(\tau),\phi(\tau)] d \tau + \frac{1}{12} \int\limits_0^t [\Phi(\tau),[\Phi(\tau),\phi(\tau)]] d\tau + \ldots 
\end{aligned}\]
 Next, this expression gets re--iterated: the quantities $\Phi(\tau)$ in the right--hand side are to be replaced, in lowest orders, with their expressions in terms of $\phi$ by this same formula. For example, to get quadratic contribution, we replace $\Phi(\tau)$ in the second term with its first--order expansion $\int \limits_0^\tau \phi(\tau_1) d \tau_1$: 
\[\begin{aligned}
 \Phi(t) = \int\limits_0^t \phi(\tau) d\tau + \frac{1}{2} \int\limits_0^t [\int \limits_0^\tau \phi(\tau_1) d\tau_1, \phi(\tau)] d\tau + \ldots =\\
 = \int_0^t \phi(\tau) d\tau - \frac{1}{2}\int\limits_{0 < t_1 < t_2 < t} [\phi(t_2), \phi(t_1)] dt_1 dt_2 + \ldots 
\end{aligned}\footnote{Fun fact: for connections with constant coefficients $\phi^a_x$, $\phi^a_y$ and variable closed contour this second term is proportional to the area bounded by said contour: $\Phi(T_{loop}) = [\phi_x,\phi_y] S_{inside} + \ldots$. This supports our concept that commutators are quantities that couple to area, that is, components of 2--forms.}\]
 To get cubic contribution, we are to reinsert \emph{this} formula into the second term of \eqref{CBCHexpanded} \emph{and} first--order expression $\int\limits_0^\tau \phi(\tau_1) d\tau_1$ into the third term of it, etc. As each insertion increases the order of the term at least by 1, the process converges in a finite number of steps in every degree. Notice that because of the structure of the formula the second contribution came out automatically ordered by derivation.

 To derive the formula \eqref{CBCH} it is sufficient to derive a differential equation on $\Phi(t)$: 
\[ \frac{d \Phi(t)}{d t} = \frac{[\Phi(t),\cdot]}{e^{[\Phi(t),\cdot]}-1} \phi(t) \]
 Together with the initial condition $\Phi(0)=0$ this implies the integral formula \eqref{CBCH}. Now, to come up with this differential, equation we start with the characteristic property of ordered exponential: 
\[ \frac{d}{d t} e^{\Phi(t)} = \frac{d}{d t} P\exp \int\limits_0^t \phi(t) dt = \phi(t) e^\Phi(t)\]
 Now, according to a well--known identity 
\[ \frac{d}{d t} e^{\Phi(t)} = \int\limits_0^1 e^{s\Phi(t)} \frac{d \Phi}{d t} e^{(1-s)\Phi(t)} ds \]
 which is equal to 
\[ \int\limits_0^1 e^{s\Phi(t)} \frac{d \Phi}{d t} e^{(1-s)\Phi(t)} ds = \left(\int\limits_0^1 e^{s\Phi(t)} \frac{d \Phi}{d t} e^{-s\Phi(t)} ds\right)e^\Phi(t) = \left( \int\limits_0^1 e^{s[\Phi(t),\cdot]} ds \frac{d \Phi}{d t} \right) e^\Phi(t), \]
 as the expression $ e^{s \Phi(t)} X e^{-s \Phi(t)}$ represents the adjoint action of group element $e^{s \Phi(t)}$ on $X$. The last expression is 
\[ \left( \int\limits_0^1 e^{s[\Phi(t),\cdot]} ds \frac{d \Phi}{d t} \right) e^\Phi(t) = \left( \frac{e^{[\Phi(t),\cdot]} - 1}{[\Phi(t),\cdot]} \frac{d \Phi}{d t} \right) e^{\Phi(t)},\]
 which is actually a valid formula because the expression $\frac{e^{[\Phi(t),\cdot]} - 1}{[\Phi(t),\cdot]}$ is a series in positive powers of $[\Phi(t),\cdot]$ matrix--to--matrix operator, so no actual inversion is involved. With that in hand, we transform our differential equation into the shape 
\[ \frac{e^{[\Phi(t),\cdot]} - 1}{[\Phi(t),\cdot]} \frac{d \Phi}{d t} = \phi(t) \]
 By acting on both sides with the inverse matrix--to--matrix operator $\frac{[\Phi(t),\cdot]}{e^{[\Phi(t),\cdot]} - 1}$ we obtain what we desired.

 The relevance of the formula \eqref{CBCH} for our matters comes from the following line of thought. If in the plane we are given a closed contour $C$ and a connection $\phi^a$ over it, we can ask ourselves: can this connection over a contour be a restriction of a \emph{flat} one, defined over the whole domain, bounded by $C$? This question is equivalent to the following one: is there any pair of smooth (Lie algebra--valued) functions $\widetilde{\phi}_x^a$, $\widetilde{\phi}_y^b$ such that 
\[\label{CBCHdeq}\left\{ \begin{aligned}
  &\partial_x \widetilde{\phi}_y^a - \partial_y \widetilde{\phi}_x^a + [\widetilde{\phi}_x, \widetilde{\phi}_y]^a = 0\\
  &\frac{d x}{d t} \widetilde{\phi}_x^a + \frac{d y}{d t} \widetilde{\phi}_y^a = \phi^a\ \text{ on the boundary} 
\end{aligned} \right. \]
 We see that, being formulated this way, this question falls under the umbrella of general extension problem, with \newline$(0,0,\text{commutator tensor})$ as a non--abelian 2--form. The point of formula \eqref{CBCH} is that it resolves this extension problem \emph{in terms of data involved in \eqref{CBCHdeq}}: it depends on $\phi^a$ on the contour $C$ and commutator tensor on the interior of the disk bound by this contour. The series \eqref{CBCH} vanishes on $\phi$'s with a trivial holonomy over $C$ --- which is exactly the case when the answer to this extension problem is positive (a flat connection must have trivial holonomy over any contour, and this is essentially the only restriction on its boundary values). In particular, the formula \eqref{CBCH} doesn't call for additional structure to multiply $\phi^a(x)\phi^a(y)$ in different points --- it refers solely to the commutator, which is the only data needed to formulate the extension problem \eqref{CBCHdeq}. No matrix multiplication is needed, and in particular, the formula \eqref{CBCH} does not depend on representation.

 A curious caveat related to the formula \eqref{CBCH} is that, just like the ordinary BCH formula, it does not have a unique form: each term of the series can be modified using Jacobi identity to be a formally different combination of $\phi^a(\tau_i)$'s and commutator tensors. In particular, if we expand \eqref{CBCH} literally, the cubic term is 
\[ \int\limits_{0 < \tau_1 < \tau_2 < \tau_3 < 1} \left\{\frac{1}{4}[[\phi(t_1),\phi(t_2)],\phi(t_3)]] + \frac{1}{12} [\phi(\tau_1),[\phi(\tau_2),\phi(\tau_3)]] \frac{1}{12} [\phi(\tau_3),[\phi(\tau_2),\phi(\tau_1)]]\right\} d\tau_1 d\tau_2 d\tau_3\]
 as opposed to more convincingly looking 
\[ \frac{1}{6} \int\limits_{0 < \tau_1 < \tau_2 < \tau_3 < 1} \left\{[[\phi(t_1),\phi(t_2)],\phi(t_3)]] + [\phi(\tau_1),[\phi(\tau_2),\phi(\tau_3)]]\right\}d\tau_1 d\tau_2 d\tau_3\]
 which coincides with the preceding given Jacobi identity. Those forms of \eqref{CBCH} are all equivalent for classical connections, but become different for general non--abelian 2--forms, which are no longer required to obey Jacobi identity. We see that, to proceed to the general case, from all the possible forms of expression \eqref{CBCH} the right one has to be selected. We leave this question for future work.

\section{Geometry of 2--dimensional integration: anchored bundles}\label{geom}

In this section, we briefly comment on differential--geometrical objects that have gauge transformations \eqref{gauge} as automorphism group and, as such, allowing for a somewhat more invariant understanding of the concepts involved in our constructions. Suppose we are studying integration over the manifold $M$; then the first necessary notion is one of so--called \emph{anchored bundles}. These are vector bundles $E \hookrightarrow A \rightarrow M$ with fixed surjective mapping (called \emph{anchor}) to the tangent bundle of $M$: $\rho: E \rightarrow TM$. This means the transition functions of this bundle can be written in block--triangular form:
\[ \label{transition} G = \begin{pmatrix} J_{\mu}^{\nu} & 0 \\
 \alpha_\mu^a & g_b^a \end{pmatrix} \]
where $J_{\alpha\beta}$ is the transition function for the tangent bundle (the Jacobian matrix of coordinate transformations). As can be seen, the bundle itself has necessarily bigger rank than $TM$, hence, some vectors on fibers project to zero under $\rho$. This kernel subbundle will be denoted as $K = \ker \rho$.

Because of our will to keep track of this $\rho$ map, it is convenient to choose the basis in fibers of $A$ in such a way that
\begin{itemize}
\item the first $d = \dim M$ basis vectors $e_1, e_2, \ldots, e_\mu, \ldots, e_d$ cover basis vectors for $TM$ $v_1 = \partial/\partial x^1, v_2 = \partial/\partial x^2, \ldots, v_\mu = \partial/\partial x^\mu$: $\rho(e_\mu) = v_\mu$;
\item the last $N = \mathrm{rk}\, K$ vectors $\widetilde{e}_{d+1}, \widetilde{e}_{d+2}, \ldots, e_a, \ldots, \widetilde{e_{d+N}}$ span kernel subbundle: $\rho(\tilde{e}_a) = 0$.
\end{itemize}
In particular, the transition functions (\ref{transition}) were written in a basis system of this kind.

It is important to notice that even if coordinates (hence, basis vectors on $TM$) are fixed, there are more than one basis on $A$ consistent with this fixation. Not only can we change the basis in the kernel subbundle, but it is also possible to shift basis vectors $e_1,\ldots,e_\mu$ along the kernel direction:
\[\begin{aligned}
 & e_\mu \mapsto e_\mu + \alpha_\mu^a \widetilde{e}_a \\
 & \widetilde{e}_a \mapsto g_a^b \widetilde{e}_b 
\end{aligned} \]
Those constitute the \emph{gauge transformations} of an anchored bundle as advertised.

Consider a section $X$ of $A$. Its basis expansion can have the form $X = v^\mu e_\mu + f^a \widetilde{e}_a$ or $\begin{pmatrix} v^\mu \\ f^a \end{pmatrix}$. Here $v^\mu$ can be identified with components of vector field $\rho{X} = v^\mu\, \partial/\partial x^\mu$. On those components gauge transformations act like
\[\begin{aligned}
 & v^\mu \mapsto v^\mu \\
 & f^a \mapsto (g^{-1})^a_b f^b + \alpha_\mu^a v^\mu
\end{aligned} \]
As $v^\mu$ part of $X$ does not transform, it could be tempting to forget it and consider $f^a$ part separately; but it would be a bad idea, as $f^a$ transformation depends on $v^\mu$ part: any specific way to consider $f$ independently of $v$ would not be invariant with respect to gauge transformation, which is, of course, undesirable.

So how do we interpret sections of $A$? In some sense, they are ``extended vectors''. While the vector $v$ corresponding to a given section $X$ is perfectly well--defined, it is impossible to define a section $X$ ``consisting of $v$ alone'': if $v$ is non-zero, we cannot invariantly set $f$ to be zero, as even if $f^a = 0$ in one basis, it will acquire non--zero components in another one, after a gauge transformation. Only if $X$ projects to zero the components $f^a$ transform through each other under gauge transformation. On the other hand, the components of $v^\mu$ can be invariantely put to zero in any circumstances.

It is good to have sections of $A$ figured out, but what if we still want to speak about sections of $A$ corresponding to given vector fields? This is the point where our much discussed splittings make their appearance. In the context of anchored bundles splittings are linear maps $\phi: TM \rightarrow A$ such that 
\[\rho \circ \phi = \mathrm{id}_{TM}.\]
 Such maps are said to split the short exact sequence $0 \rightarrow K \rightarrow A \rightarrow TM \rightarrow 0$, which is the reason for the name ``splittings''. Every splitting $\phi$ maps a vector field $v$ to a section $\phi(v)$ covering $v$: $\rho(\phi(v)) = v$. In coordinates, splittings are given by columns of matrices $\begin{pmatrix} \delta_\mu^\nu \\ \phi_\mu^\nu\end{pmatrix}$ and transform as $\phi_\mu^a \mapsto \alpha_\mu^a + g^a_b \phi_\mu^b$ Notice that the condition on splitting is linear inhomogeneous, which means that a sum of two splittings is not a splitting: $\rho \circ (\phi_1 + \phi_2) = 2 \mathrm{id}_{TM} \neq \mathrm{id}_{TM}$. Only linear combinations of the kind $ f(x^\mu) \phi_1 + (1 - f(x^\mu)) \phi_2$ can be taken without breaking the splitting condition (not unlike connections, which is not a coincidence). Another important notice is that there are many splittings of $A$, among which there is no preferrable of any kind: one can always add $\ker$--valued differential form ($\begin{pmatrix} 0 \\ \chi_\mu^a \end{pmatrix}$) to a given splitting $\phi$ and obtain another one $\phi^\prime = \phi + \chi$. No invariant criterion can select a splitting as ``initial'' or ``zero'' one.

Now that we have unraveled the geometric meaning of splittings, what about non--abelian 2--forms? As for the classical connections, it is not the best choice to give a \textit{purely} geometrical definition first\footnote{recall that the usual explanation for transition functions of connections requires treating them as differential operators, not sections of a specially constructed secondary bundle --- i.e., in a more algebraic fashion than geometric}. So instead we will give an analog of ``connections are quantities that locally are operator--valued one--forms, but with special inhomogeneous transition functions'' description. In this light, a non--abelian 2--form locally looks like a skew--symmetric bilinear map from anchored bundle to kernel subbundle $\omega: \wedge^2 A \rightarrow E$; the transition functions are again linear inhomogeneous and can be read from \eqref{gauge2f}. In a moment, we will repeat them for the sake of convenience. As indices of $A$ consist of two parts, $\omega$ data are actually threefold: 
\[(\omega_{ij}X^i Y^j)^a = \omega_{\mu\nu} X^\mu Y^\nu + \omega^a_{\mu a} X^\mu Y^a - \omega^a_{\mu a}\footnote{it is convenient to denote $\omega^a_{b\mu} = - \omega^a_{\mu b}$} Y^\mu X^a + \omega^a_{bc} X^b Y^c\]
 where $X, Y$ are sections of $A$, $X^\mu = \rho(X) \in TM$ (resp. $Y^\nu$) and $X^a, Y^b \in E$. The quantities $\omega_{\mu\nu}^a$, $\omega_{\mu b}^a$, $\omega_{bc}^a$ are precisely $A_{\mu\nu}^a$, $B^a_{\mu b}$, $C^a_{bc}$ pieces of data that have been discussed throughout this paper. So their gauge transformations (and transition functions) are given by familiar formulas 
\[ \begin{aligned}
  & \omega^a_{bc} \mapsto (g^{-1})^a_{p} g^q_b g^r_c\, \omega^{p}_{qr}\\
  & \omega^{a}_{\mu b} \mapsto (g^{-1})^a_{p} \left( g^q_b\, \omega^{p}_{\mu q} + \alpha^q_\mu g^r_b\, \omega^p_{qr} + \partial_\mu g^{p}_b\right)\\
  & \omega^{a}_{\mu \nu} \mapsto (g^{-1})^a_{p} \left( \omega^p_{\mu\nu} + \alpha^q_\mu\, \omega^p_{\mu q} - \alpha^q_\nu\, \omega^p_{\nu q} + \alpha^p_\mu \alpha^q_{\nu}\, \omega^p_{qr} + \partial_\mu \alpha_\nu^p - \partial_\nu \alpha^p_\mu \right) 
\end{aligned} \]
As previously discussed, only the $\omega^a_{bc}$ part can be invariantely put to zero, but not $\omega^a_{\mu b}$ and $\omega^a_{\mu\nu}$. If you look carefully, you'll notice that for $\omega$, as opposed to sections $\phi^i$ the $a$-- and $\mu$--type indices switch their roles in terms of the possibility of putting respective components to zero invariantely. This happens because for $\omega$ they are lower indices, while for $\phi$ they are upper: subspaces become quotients when dualizing. The last comment to be made in this section is a yet undiscussed way to derive gauge transform rules for $\omega$. It is, in a certain sense, dual to the invariance of $\omega$--exterior differential $ (\mathrm{d}_\omega \phi)^a_{\mu\nu} = \partial_\mu \phi_\nu^a - \partial_\nu \phi^a_\mu + \omega_{ij}^a \phi^i_\mu \phi^j_\nu$. Namely, we can require the following bracket: 
\[\label{bracket} \left\{ \begin{pmatrix} v^\mu \\
 a^a\end{pmatrix}, \begin{pmatrix} u^\nu\\
 b^b\end{pmatrix}\right\}_\omega = \begin{pmatrix} v^\nu \partial_\nu u^\mu - u^\nu \partial_\nu v^\mu \\
 v^\nu \partial_\nu b^a - u^\nu \partial_\mu a^a + \omega_{ij}^a \begin{pmatrix} v \\
 a\end{pmatrix}^i \begin{pmatrix} u \\
 b \end{pmatrix}^j \end{pmatrix}\]
to be invariant under gauge transformations: if $\omega$ transforms as stated, this bracket calculated in either initial or final basis system yields covariantly--related results. It extends the usual Lie bracket of vector fields $\{v,u\}^\mu = v^\nu \partial_\nu u^\mu - u^\nu \partial_\nu v^\mu$: $\rho(\{X,Y\}_{\omega}) = \{\rho(X),\rho(Y)\}$. Broadly speaking, the special properties of tangent bundles (like torsion spaces, for example) stem from the structure of Lie bracket on them; a 2--form lifts this structure to an anchored bundle in question ``in dimension two'' (because it defines a bilinear operation), while no ``integrability in dimension three'' is guaranteed. Making this statement precise is one of the important questions for further investigation in the field of non--abelian 2--integrals.

\section{Conclusion}\label{conclusion}

In this work, we have outlined the foundations of future non--abelian 2--dimensional integration theory, which should extend the theory of connections on vector bundles. The key objects of this theory are splittings $\phi^a_\mu$ and non--abelian 2--forms $(A_{\mu\nu}^a, B_{\mu b}^a, C_{bc}^a)$, which are precise analogs of vector bundle sections $v^\alpha$ and classical connections $A_{\mu \beta}^\alpha$ in one--dimensional theory. At the heart of the theory is the differential relation
\[ \partial_\mu \phi^a_\nu - \partial_\nu \phi^a_\mu + A_{\mu\nu}^a + B_{\mu b}^a \phi^b_\nu - B_{\nu b}^a \phi^b_\mu + C^a_{bc} \phi^b_{\mu} \phi^c_{\nu} = 0\]
that generalizes the parallel transport equation
\[ \partial_\mu v^\alpha = A_{\mu \beta}^\alpha v^\beta\]
This equation admits a group of gauge symmetries $ \phi^a_\mu(x) \mapsto g^a_b (x) \phi^b_\mu (x) + \alpha^a_\mu(x)$, paired with appropriate changes in $(A,B,C)$ (which in particular mix those components with each other). Splittings $\phi^a_\mu$ that obey this relation for given $\omega$ are called $\omega$--flat. When restricted onto a two--dimensional surface, they always exist: in fact, one can take arbitrary splitting on an interval on $Ox$ axis and then ``spread'' it in $\omega$--flat way over a 2--dimensional domain. This spreading is defined by a deformation of the contour $I \rightarrow I_t,\ t \in [0;1]$, that is, by a 2--homotopy moving it: the contour sweeps the two--dimensional domain along the way. Changes in $\phi^a_x$ along the way depend on additional data: one has to fix $\phi_y^a$ in transversal direction to have $\omega$--flatness condition fix $\phi_x^a$ unambigously. The whole process is a 2--dimensional analog of parallel transport along 1--paths. So we see that 2--dimensional flatness equations are too weak to determine parallel transport of $\phi_x^a$ all by themselves, but they allow $\phi^a_y$ components to transport $\phi^a_x$ and vice versa. Familiar examples of flatness relations are exterior differentials ($a$ runs over one value, $B^a_{\mu b} = 0$, $C^a_{bc} = 0$):
\[ \partial_\mu \phi_\nu - \partial_\nu \phi_\mu + A_{\mu\nu} = 0\]
and flatness condition for connections ($A^a_{\mu\nu} = 0$, $B^a_{\mu b} = 0$, $C^a_{bc}\phi^b_\mu \phi^c_\nu = [\phi_\mu,\phi_\nu]^a$ in Lie algebra):
\[ \partial_\mu \phi^a_\nu - \partial_\nu\phi^a_\mu + [\phi_\mu,\phi_\nu]^a = 0 \]
Both gauge transformation rules and relations between splittings and 2--forms can be given a geometric interpretation in the framework of anchored bundles.

To move from parallel transport to integrals, we suggested defining the latter as obstructions to parallel--transporting a closed contour through a region it bounds to a point. The problem is, if we try to transport a general splitting on a circle by a contracting homotopy, this splitting will develop a singularity at the limiting point. Only for some initial splittings the singular terms can be cancelled by a proper choice of $\phi^a_\text{transversal}$; an integral is just a functional that selects contractible boundary splittings among all. In other words, a 2--dimensional non--abelian integral is a functional $I[\omega,\phi^{(0)}]$ that vanishes whenever the system 
\[\left\{ \begin{aligned}
  & \partial_\mu \widetilde{\phi}^a_\nu - \partial_\nu \widetilde{\phi}^a_\mu + A_{\mu\nu}^a + B_{\mu b}^a \widetilde{\phi}^b_\nu - B_{\nu b}^a \widetilde{\phi}^b_\mu + C^a_{bc} \widetilde{\phi}^b_{\mu} \widetilde{\phi}^c_{\nu} = 0\\
  & \widetilde{\phi}^a_{\mu} dx^\mu = \phi^{a(0)} dt\ \text{ on the boundary} 
\end{aligned}\right.\]
posesses a smooth $\widetilde{\phi}$ solution. As perverse as this definition is, it incorporates both abelian 2--integrals (with $I[\omega,\phi^{0}]$ being Stokes functional $\int \omega + \oint \phi^{(0)}$) and flat connections (with $I = \log P\exp \oint \phi^{0}$ as a function of $\phi^{(0)}$ and commutator tensor). The specifics of the 2--dimensional case (as opposed to the 1--dimensional one) requires $I$ to be a functional in boundary conditions as opposed to a finite--dimensional matrix. To support our claim that the solvability of said system is controlled by a single vector--valued functional, we presented a very basic parameter--counting argument (reproducing, however, the abelian case correctly), based on a combination of Fourier and Taylor expansions of splitting coefficients, and reproduced the derivation of the continuous-limit Baker--Campbell--Hausdorff formula, expressing the necessary quantity in the case of flat classical connections in terms of boundary values and commutators. Obtaining a closed integral formula for non--abelian 2--integrals in general case that interpolates between the two extremes is an exciting and challenging problem we are going to address in further research. A promising starting point can be, for example, finding a BCH analog of the non--abelian Stokes formula\cite{Arefeva:1979dp} that solves the extension problem for $\omega = (\mathrm{Curv}(A)_{\mu\nu}^a,0,f^a_{bc})$. Besides extending known closed formulas for integrals to new classes of non--abelian 2--forms, there are a variety of questions left open by this research for future investigation. For example, the intrinsic needs of non--abelian integration theory naturally require figuring out
\begin{itemize}
\item given integrals $I_1$, $I_2$ of one and the same 2--form over two adjacent domains $D_1$, $D_2$ sharing a segment of their boundary, how to express the integral over $D_1 \cup D_2$ in terms of $I_1$ and $I_2$? what is the analog of the multiplicative property of $P$--exponentials?
\item how do 2--dimensional integrals change when their contour is deformed in some ambient 3--dimensional space? this question involves coming up with the formula for the exterior differential of non--abelian 2--forms
\item can we somehow construct a non--abelian 2--form from two classical connections? i.e., what does non--abelian wedge product for the case of dimensions 1 and 1 look like?
\end{itemize}
Meanwhile, the exploration of extrinsic questions targeted at applications of non--abelian integration can initially be led along the lines sketched in the introduction, which include clarifying the relations of non--abelian 2--forms with two--dimensional quantum field theory, figuring out (after exterior differentiation has been settled) the properties of corresponding Yang--Mills type fields and strings coupled to 2--form fields, hunting for characters and Chern--Weil type homomorphisms in 2--connection case, describing (again, after exterior differentiation) the moduli spaces of flat (closed) 2--forms, etc., etc. This list is by no means thought to be complete, and we fervently hope these questions become more tackleable as more effort is put into the development of non--abelian 2--form theory.

\section{Acknowledgements}

I am extensively grateful to Alexei Morozov for fruitful discussions and valuable advice, and to the participants of MIPT/ITEP theoretical and mathematical physics research seminar for their patience and enthusiasm. I am also massively indebted to the anomynous referee, who suggested looking at the approach described in section \ref{assault2} that turned out to be rather fruitful. The work was supported by Russian Science Foundation, grant \textnumero 20--71--10073.

\bibliographystyle{utphys}\bibliography{2-conn-complete-v0.0.2}

\end{document}